\documentclass[preprintnumbers,amsmath,amssymb,floatfix,11pt,prd,onecolumn,showpacs,
superscriptaddress,nofootinbib]{revtex4}
\usepackage{graphicx}
\usepackage{epsfig}
\usepackage{bm}
\usepackage{amsfonts}

\begin{document}

\title{\large{\bf DYNAMICS OF INTERACTING GENERALIZED COSMIC CHAPLYGIN GAS IN BRANE-WORLD SCENARIO}}

\author{ \textbf{Prabir
Rudra}}\email{prudra.math@gmail.com} \affiliation{Department of
Mathematics, Bengal Engineering and Science University, Shibpur,
Howrah-711 103, India.}

\begin{abstract}

\noindent

In this work we explore the background dynamics when dark energy
is coupled to dark matter with a suitable interaction in the
universe described by brane cosmology. Here DGP and the RSII brane
models have been considered separately. Dark energy in the form of
Generalized Cosmic Chaplygin gas is considered. A suitable
interaction between dark energy and dark matter is considered in
order to at least alleviate (if not solve) the cosmic coincidence
problem. The dynamical system of equations is solved numerically
and a stable scaling solution is obtained. A significant attempt
towards the solution of the cosmic coincidence problem is taken.
The statefinder parameters are also calculated to classify the
dark energy models. Graphs and phase diagrams are drawn to study
the variations of these parameters. It is also seen that the
background dynamics of Generalized Cosmic Chaplygin gas is
consistent with the late cosmic acceleration, but not without
satisfying certain conditions. It has been shown that the universe
in both the models follows the power law form of expansion around
the critical point, which is consistent with the known results.
Future singularities were studied and our models were declared
totally free from any types of such singularities. Finally, some
cosmographic parameters were also briefly studied. Our
investigation led to the fact that although GCCG with a far lesser
negative pressure compared to other DE models, can overcome the
relatively weaker gravity of RS II brane, with the help of the
negative brane tension, yet for the DGP brane model with much
higher gravitation, the incompetency of GCCG is exposed, and it
cannot produce the accelerating scenario until it reaches the
phantom era.

\end{abstract}

\maketitle

\newpage

\section{INTRODUCTION}

\noindent

Recent cosmological observations have indicated that the
observable universe enters into an epoch of accelerated expansion
\cite{Perlmutter1, Spergel1}. In the quest of finding a suitable
model for universe, Cosmologists started to investigate the root
cause that is triggering this expansion. Within the framework of
the general relativity, the acceleration can be phenomenally
attributed to the existence of a mysterious negative pressure
component which violates the strong energy condition i.e.
$\rho+3p<0$. Because of its invisible nature this energy component
is aptly termed as dark energy (DE) \cite{Riess1}. Moreover, quite
surprisingly, observations spilled out definitively that about 70
percent of the Universe is filled by this unknown ingredient and
in addition that about 25 percent of this is composed by dark
matter(DM).

\noindent

With the introduction of DE, search began for different candidates
that can effectively play its role.  DE represented by a scalar
field \footnote{ in the presence of a scalar field the transition
from a universe filled with matter to an exponentially expanding
universe is justified } \cite{Nojiri1} is often called
quintessence. Not only scalar field but also there are other Dark
fluid models like Chaplygin gas which plays the role of DE very
efficiently. Extensive research saw Chaplygin gas (CG)
\cite{Kamenshchik1, Gorini1}, get modified into Generalized
Chaplygin gas (GCG) \cite{Gorini2, Alam1, Bento1, Carturan1,
Barreiro1} and then to Modified Chaplygin gas (MCG)
\cite{Benaoum1, Debnath1}. In this context it is worth mentioning
that Interacting MCG in Loop quantum cosmology (LQC) was studied
by Jamil et al \cite{Jamil1}. Dynamics of MCG in Braneworld was
studied by Rudra et al \cite{Rudra1}. Other than these other forms
of Chaplygin gas models have also been proposed such as Variable
Modified Chaplygin gas (VMCG) \cite{Debnath2} and New Variable
Modified Chaplygin gas (NVMCG) \cite{Chakraborty1}. Other existing
forms of DE are phantom \cite{Nojiri2}, k-essence \cite{Bamba1},
tachyonic field \cite{Nojiri3}, etc.

\noindent

In 2003, P. F. Gonz´alez-Diaz \cite{Gonz´alez-Diaz1} introduced
the generalized cosmic Chaplygin gas (GCCG) model. The speciality
of the model being that it can be made stable and free from
unphysical behaviours even when the vacuum fluid satisfies the
phantom energy condition. In the previous studies related to DE
corresponding to phantom era Big-Rip is essential, as the time
gradient of scale-factor blows to infinity in finite time. For the
first time P. F. Gonz´alez-Diaz including the GCCG model showed
that Big Rip, i.e., singularity at a finite time is totally out of
question. Hence in such models there is no requirement for
evaporation of black hole to zero mass. The Equation of state
(EoS) of the GCCG model is
\begin{equation}
p=-\rho^{-\alpha}\left[C+\left\{\rho^{(1+\alpha)}-C\right\}^{-\omega}\right]
\end{equation}
where $C=\frac{A}{1+\omega}-1$, with A being a constant that can
take on both positive and negative values, and $-{\cal
L}>\omega>0$, ${\cal L}$ being a positive definite constant, which
can take on values larger than unity. GCCG can explain the
evolution of the universe starting from the dust era to $\Lambda
CDM$, radiation era, matter dominated quintessence and lastly
phantom era \cite{Chakraborty2}. In this context it should be
stated that in \cite{Chowdhury1} Chowdhury and Rudra studied
Interacting GCCG in Loop Quantum Cosmology (LQC).

\noindent

Currently, we live in a special epoch where the densities of DE
and DM are comparable. Although they have evolved independently
from different mass scales. This is known as the famous cosmic
coincidence problem. till date several attempts have been made to
find a solution to this problem \cite{del1, Leon1, Jimenez1,
Berger1, Zhang1, Griest1, Jam1, Jam2, Jam3, Jam4, Jam5}. A
suitable interaction between DE and DM is required if we wish to
find an effective solution to this problem. It is obvious that
there has been a transition from a matter dominated universe to
dark energy dominated universe, by exchange of energy at an
appropriate rate. Now the expansion history of the universe as
determined by the supernovae and CMB data \cite{Jam2, Jam3} bounds
us to fix the decay rate such that it is proportional to the
present day Hubble parameter. Keeping the fact in mind
cosmologists all over the world have studied and proposed a
variety of interacting DE models \cite{Set1, Set2, Hu1, Wu1, Jam6,
Jam7, Dalal1}.

\noindent

As we have stated earlier, modifying the right hand side of
Einstein's equation (DE approach) was not the only way to explain
the increase in the rate of the expansion. We can also modify the
gravity part of the left hand side in order to demonstrate the
present day universe. In this context Brane-gravity was introduced
and brane cosmology was developed. A review on brane-gravity and
its various applications with special attention to cosmology is
available in \cite{Rub1, Maartens1, Brax1, Csa1}. In this work we
consider the two most popular brane models, namely DGP and RS II
branes. Our main aim of this work is to examine the nature of the
different physical parameters for the universe around the stable
critical points in two brane world models in presence of GCCG.
Impact of any future singularity caused by the DE in brane world
models will be studied. Finally some cosmographic parameters will
be studied in brief.

\noindent

This paper is organized as follows: Section 2 comprises of the
analysis in RS II brane model. Section 3 deals with the analysis
in DGP brane model. In section 4, a detailed graphical analysis
for the phase plane is done. In section 5, future singularities
arising from our models are studied, followed by the study of some
cosmographic parameters in section 6. Finally the paper ends with
some concluding remarks in section 7.

\section{MODEL 1: RS II BRANE MODEL}

\noindent

Randall and Sundrum \cite{rs1, rs2} proposed a bulk-brane model to
explain the higher dimensional theory, popularly known as RS II
brane model. According to this model we live in a four dimensional
world (called 3-brane, a domain wall) which is embedded in a 5D
space time (bulk). All matter fields are confined in the brane
whereas gravity can only propagate in the bulk. The consistency of
this brane model with the expanding universe has given popularity
to this model of late in the field of cosmology.

\noindent

In RS II model the effective equations of motion on the 3-brane
embedded in 5D bulk having $Z_{2}$-symmetry are given by
\cite{Maartens1, Maartens2, rs2, Shiromizu1, Maeda1, Sasaki1}
\begin{equation}
^{(4)}G_{\mu\nu}=-\Lambda_{4}q_{\mu\nu}+\kappa^{2}_{4}\tau_{\mu\nu}+\kappa^{4}_{5}\Pi_{\mu\nu}-E_{\mu\nu}
\end{equation}
where

\begin{equation}
\kappa^{2}_{4}=\frac{1}{6}~\lambda\kappa^{4}_{5}~,
\end{equation}
\begin{equation}
\Lambda_{4}=\frac{1}{2}~\kappa^{2}_{5}\left(\Lambda_{5}+\frac{1}{6}~\kappa^{2}_{5}\lambda^{2}\right)
\end{equation}
and
\begin{equation}
\Pi_{\mu\nu}=-\frac{1}{4}~\tau_{\mu\alpha}\tau^{\alpha}_{\nu}+\frac{1}{12}~\tau\tau_{\mu\nu}+\frac{1}{8}~
q_{\mu\nu}\tau_{\alpha\beta}\tau^{\alpha\beta}-\frac{1}{24}~q_{\mu\nu}\tau^{2}
\end{equation}
and $E_{\mu\nu}$ is the electric part of the 5D Weyl tensor. Here
$\kappa_{5},~\Lambda_{5},~\tau_{\mu\nu}$ and $\Lambda_{4}$ are
respectively the 5D gravitational coupling constant, 5D
cosmological constant, the brane tension (vacuum energy), brane
energy-momentum tensor and effective 4D cosmological constant. The
explicit form of the above modified Einstein equations in flat
universe are

\begin{equation}
3H^{2}=\Lambda_{4}+\kappa^{2}_{4}\rho+\frac{\kappa^{2}_{4}}{2\lambda}~\rho^{2}+\frac{6}{\lambda
\kappa^{2}_{4}}\cal{U}
\end{equation}
and
\begin{equation}
2\dot{H}+3H^{2}=\Lambda_{4}-\kappa^{2}_{4}p-\frac{\kappa^{2}_{4}}{2\lambda}~\rho
p-\frac{\kappa^{2}_{4}}{2\lambda}~\rho^{2}-\frac{2}{\lambda
\kappa^{2}_{4}}\cal{U}
\end{equation}
The dark radiation $\cal{U}$ obeys

\begin{equation}
\dot{\cal U}+4H{\cal U}=0
\end{equation}
where $\rho=\rho_{gccg}+\rho_{m}$ and $p=p_{gccg}+p_{m}$ are the
total energy density and pressure respectively.

\noindent

As in the present problem the interaction between DE and
pressureless DM has been taken into account for interacting DE and
DM the energy balance equation will be
\begin{equation}
\dot{\rho}_{gccg}+3H\left(1+\omega_{gccg}\right)\rho_{gccg}=-Q,~~~for
~GCCG~ and ~
\end{equation}
\begin{equation}
\dot{\rho}_m+3H\rho_m=Q, ~for~ the~ DM~ interacting ~with~ GCCG.
\end{equation}
where $Q=3bH\rho$ is the interaction term, $b$ is the coupling
parameter (or transfer strength) and $\rho=\rho_{gccg}+\rho_m$ is
the total cosmic energy density which satisfies the energy
conservation equation $\dot{\rho}+3H\left(\rho+p\right)=0$
\cite{Guo1, del1}.

\noindent

As we lack information about the fact, how does DE and DM interact
so we are not able to estimate the interaction term from the first
principles. However, the negativity of $Q$ immediately implies the
possibility of having negative DE in the early universe which is
overruled by to the necessity of the second law of thermodynamics
to be held \cite{Alcaniz1}. Hence $Q$ must be positive and small.
From the observational data of 182 Gold type Ia supernova samples,
CMB data from the three year WMAP survey and the baryonic acoustic
oscillations from the Sloan Digital Sky Survey, it is estimated
that the coupling parameter between DM and DE must be a small
positive value (of the order of unity), which satisfies the
requirement for solving the cosmic coincidence problem and the
second law of thermodynamics \cite{Feng1}. Due to the underlying
interaction, the beginning of the accelerated expansion is shifted
to higher redshifts. The continuity equations for dark energy and
dark matter are given in equations (9) and (10). Now we shall
study the dynamical system assuming $\Lambda_{4}={\cal U}=0$ (in
absence of cosmological constant and dark radiation).

\subsection{DYNAMICAL SYSTEM ANALYSIS}

\noindent

Due to the complexity of the equations, it is very difficult to
find direct solutions for this system. So in order to avoid these
complex calculations, we undertake the dynamical system analysis
for our further evaluations. In this subsection we plan to analyze
the dynamical system. Before proceeding, the physical parameters
are converted into some dimensionless form, given by
\begin{equation}
x=\ln a, ~~~~~~~ u=\frac{\rho_{gccg}}{3H^2}, ~~~~~~~
v=\frac{\rho_m}{3H^2}
\end{equation}
where the present value of the scale factor $a_0=1$ is assumed.
Now using equations (1), (6), (7), (9), (10) and (11) we get,

$$\frac{du}{dx}=-3b\left(u+v\right)+3u-3u\omega_{gccg}^{RS II}-6\kappa_{4}^{2}u\left(u+v\right)+\frac{9\kappa_{4}^{2}u\left(u+v\right)^{3}}{2\lambda\left(1-\kappa_{4}^{2}\left(u+v\right)\right)}$$
\begin{equation}
-\frac{9\kappa_{4}^{2}\left(u+v\right)^{3}}{4\lambda^{2}\left(1-\kappa_{4}^{2}\left(u+v\right)\right)^{2}}\left\{C+\left(\frac{4\lambda^{2}\left(1-\kappa_{4}^{2}\left(u+v\right)\right)^{2}u^{2}}{\kappa_{4}^{4}\left(u+v\right)^{4}}-C\right)^{-w}\right\}
\end{equation}

\begin{equation}
\frac{dv}{dx}=3b\left(u+v\right)+3v-3\kappa_{4}^{2}v\left(u+v\right)-\frac{3\kappa_{4}^{4}v\left(u+v\right)^{3}}{4\lambda^{2}\left(1-\kappa_{4}^{2}\left(u+v\right)\right)^{2}u}\left\{C+\left(\frac{4\lambda^{2}\left(1-\kappa_{4}^{2}\left(u+v\right)\right)^{2}u^{2}}{\kappa_{4}^{4}\left(u+v\right)^{4}}-C\right)^{-w}\right\}
\end{equation}

Where, $\omega_{gccg}^{RS II}$ is the EoS parameter for GCCG in RS
II brane determined as

\begin{equation}
\omega_{gccg}^{RS
II}=\frac{p_{gccg}}{\rho_{gccg}}=-\frac{\kappa_{4}^{4}\left(u+v\right)^{4}}{4\lambda^{2}u^{2}\left(1-\kappa_{4}^{2}\left(u+v\right)\right)^{2}}\left[C+\left\{\frac{4\lambda^{2}u^{2}\left(1-\kappa_{4}^{2}\left(u+v\right)\right)^{2}-2C\kappa_{4}^{4}\left(u+v\right)^{4}}{\kappa_{4}^{4}\left(u+v\right)^{4}}\right\}^{-w}\right]
\end{equation}

\vspace{1mm}

In the above calculations for mathematical simplicity we have
considered $\alpha=1$.

\subsubsection{\bf CRITICAL POINTS}

\noindent

The critical points of the above system are obtained by putting
$\frac{du}{dx}=0=\frac{dv}{dx}$. But due to the complexity of
these equations, it is not possible to find a solution in terms of
the involved parameters. So we find a numerical solution for the
above system, by putting the following values to the different
parameters appearing in the system. We take,

$$b=1.5,~~~~~ \kappa_{4}=1,~~~~~ \lambda=0.1,~~~~~ C=-1,~~~~~ w=-1$$\\

\noindent

and get the following critical point for the above dynamical
system.\\

\begin{equation}
u_{c}=1.51586 ~~~~~~~~~~~~~ v_{c}=1.79374
\end{equation}

\noindent

The critical point correspond to the era dominated by DM and GCCG
type DE. For the critical point $(u_{c},v_{c})$, the equation of
state parameter given by equation (14) of the interacting DE takes
the form

\begin{equation}
\omega_{gccg}^{RS
II}=\frac{p_{gccg}}{\rho_{gccg}}=-\frac{\kappa_{4}^{4}\left(u_{c}+v_{c}\right)^{4}}{4\lambda^{2}u_{c}^{2}\left(1-\kappa_{4}^{2}\left(u_{c}+v_{c}\right)\right)^{2}}\left[C+\left\{\frac{4\lambda^{2}u_{c}^{2}\left(1-\kappa_{4}^{2}\left(u_{c}+v_{c}\right)\right)^{2}-2C\kappa_{4}^{4}\left(u_{c}+v_{c}\right)^{4}}{\kappa_{4}^{4}\left(u_{c}+v_{c}\right)^{4}}\right\}^{-w}\right]
\end{equation}

\vspace{2mm}

\subsubsection{\bf STABILITY AROUND CRITICAL POINT}

Now we check the stability of the dynamical system  (eqs. (12) and
(13)) about the critical point. In order to do this, we linearize
the governing equations about the critical point i.e.,
\begin{equation}
u=u_c+\delta u ~~  and ~~  v=v_c+\delta v,
\end{equation}
Now if we assume $f=\frac{du}{dx}$ and $g=\frac{dv}{dx}$, then we
may obtain
\begin{equation}
\delta\left(\frac{du}{dx}\right)=\left[\partial_{u}
f\right]_{c}\delta u+\left[\partial_{v} f\right]_{c}\delta v
\end{equation}
and
\begin{equation}
\delta\left(\frac{dv}{dx}\right)=\left[\partial_{u}
g\right]_{c}\delta u+\left[\partial_{v} g\right]_{c}\delta v
\end{equation}

where

\vspace{2mm}

\noindent

$$\partial_{u}{f}=\frac{1}{4\kappa_{4}^{2}u^{2}\left(u+v\right)^{2}\left(-1+\kappa_{4}^{2}\left(u+v\right)\right)^{3}\lambda^{2}}3\left[C\kappa_{4}^{6}\left(u+v\right)^{5}\left\{u\left(-3+\kappa_{4}^{2}u\right)+v-\kappa_{4}^{2}v^{2}\right\}\right.$$
$$\left.+2u^{2}\left\{-1+\kappa_{4}^{2}\left(u+v\right)\right\}\lambda\left\{4\kappa_{4}^{8}\left(u+v\right)^{4}\left(2u+v\right)\lambda+6u\left(u+2v\right)\lambda-2\kappa_{4}^{2}\left(u+v\right)\left(u\left(2-b+6u\right)\right.\right.\right.$$
$$\left.\left.\left.+\left(2-b+12u\right)v\right)\lambda+\kappa_{4}^{6}\left(u+v\right)^{3}\left(3\left(u+v\right)^{2}\left(3u+v\right)+2\left(\left(b-10\right)u+\left(b-6\right)v\right)\lambda\right)+\kappa_{4}^{4}\left(u+v\right)^{2}\right.\right.$$
\begin{equation}
\left.\left.\left(-3\left(u+v\right)^{2}\left(4u+v\right)+2\left(u\left(8-2b+3u\right)-2\left(b-3\left(1+u\right)\right)v\right)\lambda\right)\right\}\right]
\end{equation}

\vspace{2mm}

\noindent

$$\partial_{v}{f}=-\frac{3}{2\kappa_{4}^{2}u\left(u+v\right)^{2}\left(-1+\kappa_{4}^{2}\left(u+v\right)\right)^{3}\lambda^{2}}\left[C\kappa_{4}^{6}\left(u+v\right)^{5}\left\{-2+\kappa_{4}^{2}\left(u+v\right)\right\}+\lambda u\left\{-1+\kappa_{4}^{2}\left(u+v\right)\right\}\right.$$
$$\left.\left\{-6u^{2}\lambda+4\kappa_{4}^{8}u\left(u+v\right)^{4}\lambda+2\kappa_{4}^{2}\left(u+v\right)\left(6u^{2}+b\left(u+v\right)\right)\lambda+2\kappa_{4}^{6}\left(u+v\right)^{3}\left(3u\left(u+v\right)^{2}+\left(\left(b-4\right)u+bv\right)\lambda\right)\right.\right.$$
\begin{equation}
\left.\left.+\kappa_{4}^{4}\left(u+v\right)^{2}\left(-9u\left(u+v\right)^{2}-2\left(u\left(3u+2b-2\right)+2bv\right)\lambda\right)\right\}\right]
\end{equation}

\vspace{2mm}

\noindent

$$\partial_{u}{g}=\frac{3}{4}\left[4b-4\kappa_{4}^{2}v-\frac{8\kappa_{4}^{4}vw\left(u+v\right)^{2}\left\{u+\kappa_{4}^{2}uv+v\left(-1+\kappa_{4}^{2}v\right)\right\}\left\{-C+\frac{4u^{2}\left(-1+\kappa_{4}^{2}\left(u+v\right)\right)^{2}\lambda^{2}}{\kappa_{4}^{4}\left(u+v\right)^{4}}\right\}^{-w}}{\left\{-1+\kappa_{4}^{2}\left(u+v\right)\right\}\left\{C\kappa_{4}^{4}\left(u+v\right)^{4}-4u^{2}\left(-1+\kappa_{4}^{2}\left(u+v\right)\right)^{2}\lambda^{2}\right\}}\right.$$
$$\left.-\frac{2\kappa_{4}^{6}v\left(u+v\right)^{3}\left\{C+\left(-C+\frac{4u^{2}\left(-1+\kappa_{4}^{2}\left(u+v\right)\right)^{2}\lambda^{2}}{\kappa_{4}^{4}\left(u+v\right)^{4}}\right)^{-w}\right\}}{u\left(1-\kappa_{4}^{2}\left(u+v\right)\right)^{3}\lambda^{2}}\right.$$
$$\left.-\frac{3\kappa_{4}^{4}v\left(u+v\right)^{2}\left(C+\left(-C+\frac{4u^{2}\left(-1+\kappa_{4}^{2}\left(u+v\right)\right)^{2}\lambda^{2}}{\kappa_{4}^{4}\left(u+v\right)^{4}}\right)^{-w}\right)}{u\left(-1+\kappa_{4}^{2}\left(u+v\right)\right)^{2}\lambda^{2}}\right.$$
\begin{equation}
\left.+\frac{\kappa_{4}^{4}v\left(u+v\right)^{3}\left(C+\left(-C+\frac{4u^{2}\left(-1+\kappa_{4}^{2}\left(u+v\right)\right)^{2}\lambda^{2}}{\kappa_{4}^{4}\left(u+v\right)^{4}}\right)^{-w}\right)}{u^{2}\left(-1+\kappa_{4}^{2}\left(u+v\right)\right)^{2}\lambda^{2}}\right]
\end{equation}

\vspace{2mm}

\noindent

$$\partial_{v}{g}=\frac{3}{4}\left[4+4b-4\kappa_{4}^{2}v-4\kappa_{4}^{2}\left(u+v\right)+\frac{8\kappa_{4}^{2}uvw\left(u+v\right)^{2}\left\{\kappa_{4}^{2}\left(u+v\right)-2\right\}\left\{-C+\frac{4u^{2}\left(\left(\kappa_{4}^{2}\left(u+v\right)-1\right)\right)^{2}\lambda^{2}}{\kappa_{4}^{4}\left(u+v\right)^{4}}\right\}^{-w}}{\left\{\kappa_{4}^{2}\left(u+v\right)-1\right\}\left\{C\kappa_{4}^{4}\left(u+v\right)^{4}-4u^{2}\left(\kappa_{4}^{2}\left(u+v\right)-1\right)^{2}\lambda^{2}\right\}}\right.$$
$$\left.-\frac{2\kappa_{4}^{6}v\left(u+v\right)^{3}\left\{C+\left(-C+\frac{4u^{2}\left(\kappa_{4}^{2}\left(u+v\right)-1\right)^{2}\lambda^{2}}{\kappa_{4}^{4}\left(u+v\right)^{4}}\right)^{-w}\right\}}{u\left\{1-\kappa_{4}^{2}\left(u+v\right)\right\}^{3}\lambda^{2}}\right.$$
$$\left.-\frac{3\kappa_{4}^{4}v\left(u+v\right)^{2}\left\{C+\left(-C+\frac{4u^{2}\left(\kappa_{4}^{2}\left(u+v\right)-1\right)^{2}\lambda^{2}}{\kappa_{4}^{4}\left(u+v\right)^{4}}\right)^{-w}\right\}}{u\left\{-1+\kappa_{4}^{2}\left(u+v\right)\right\}^{2}\lambda^{2}}\right.$$
\begin{equation}
\left.-\frac{\kappa_{4}^{4}\left(u+v\right)^{3}\left\{C+\left(-C+\frac{4u^{2}\left(\kappa_{4}^{2}\left(u+v\right)-1\right)^{2}\lambda^{2}}{\kappa_{4}^{4}\left(u+v\right)^{4}}\right)^{-w}\right\}}{u\left\{-1+\kappa_{4}^{2}\left(u+v\right)\right\}^{2}\lambda^{2}}\right]
\end{equation}

\noindent

The Jacobian matrix of the above system is given by,
$$
J_{\left(u,v\right)}^{(RSII)}=\left(\begin{array}{c}\frac{\delta
f}{\delta u} ~~~~~ \frac{\delta f}{\delta v}\\ \frac{\delta
g}{\delta u}~~~~~ \frac{\delta g}{\delta v} \end{array}\right)
$$

The eigen values of the above matrix are calculated at the
critical point $(u_{c}, v_{c})$ and are found to be~~ ${\bf
\lambda_{1}=-1169.45,~~~ \lambda_{2}=-3.38251}$. Hence it is a
stable node.

\vspace{2mm}

\begin{figure}

\includegraphics[height=2in]{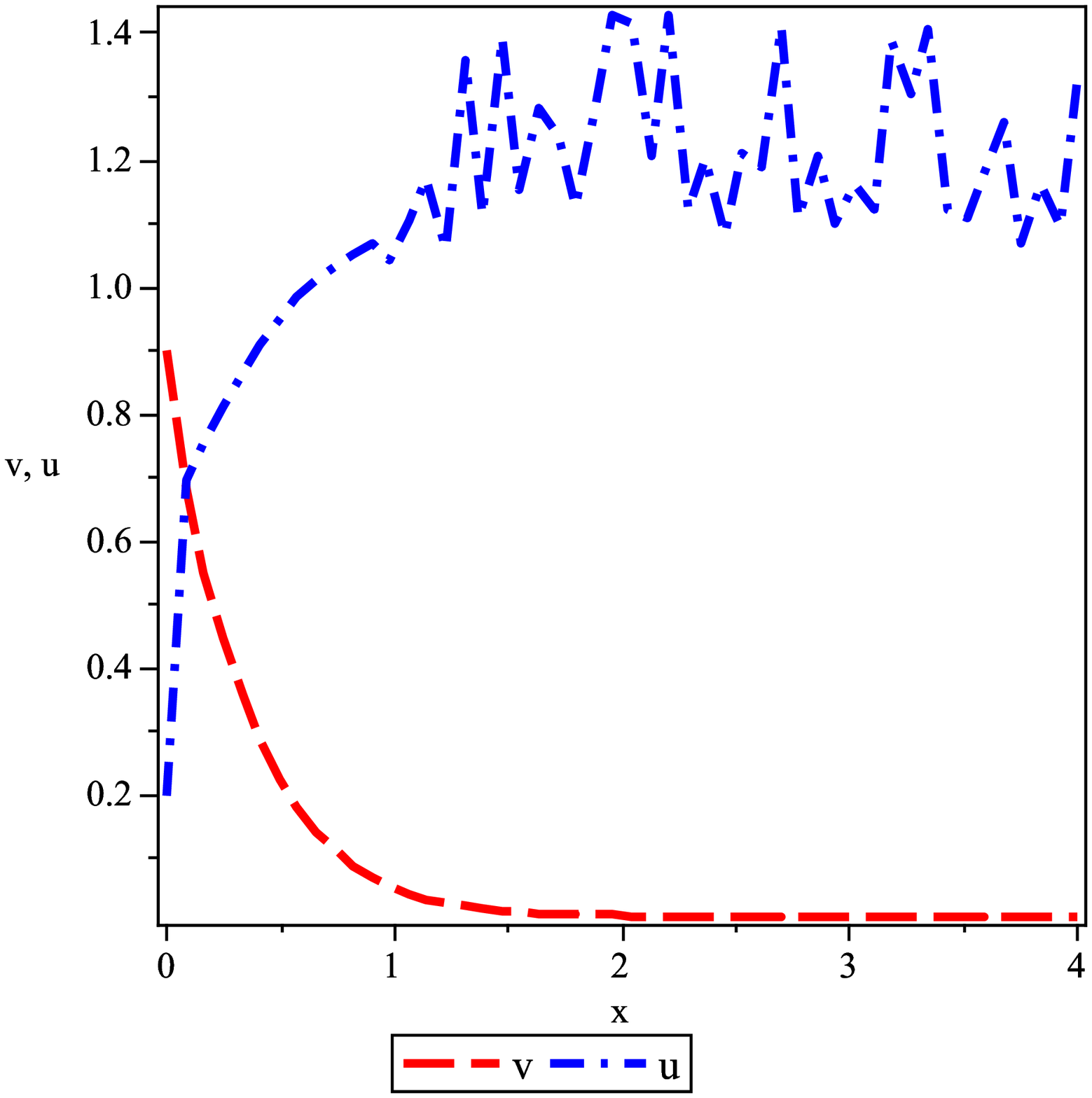}~~~~~~~~\includegraphics[height=2in]{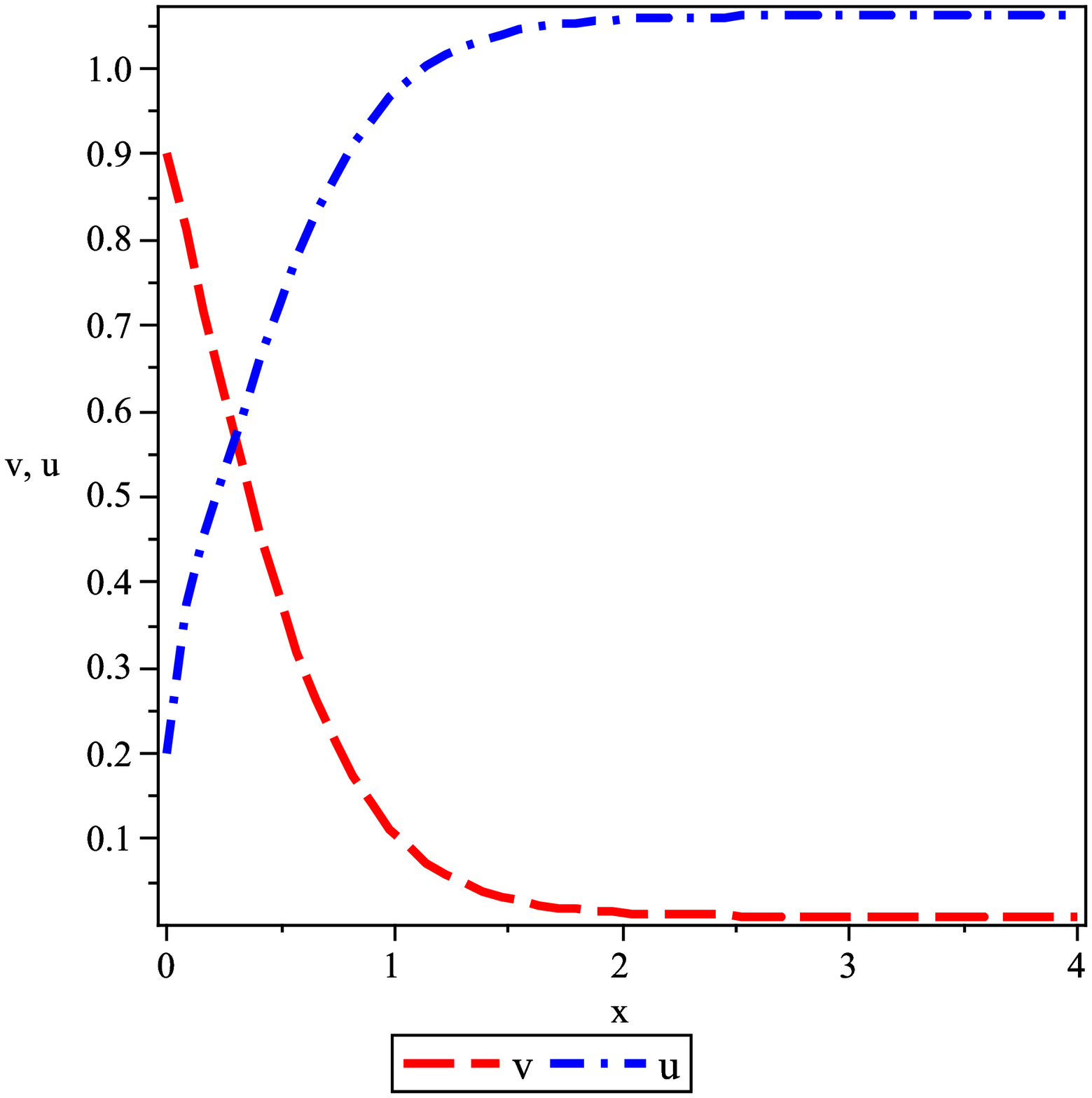}~~~~~~~~\includegraphics[height=2in]{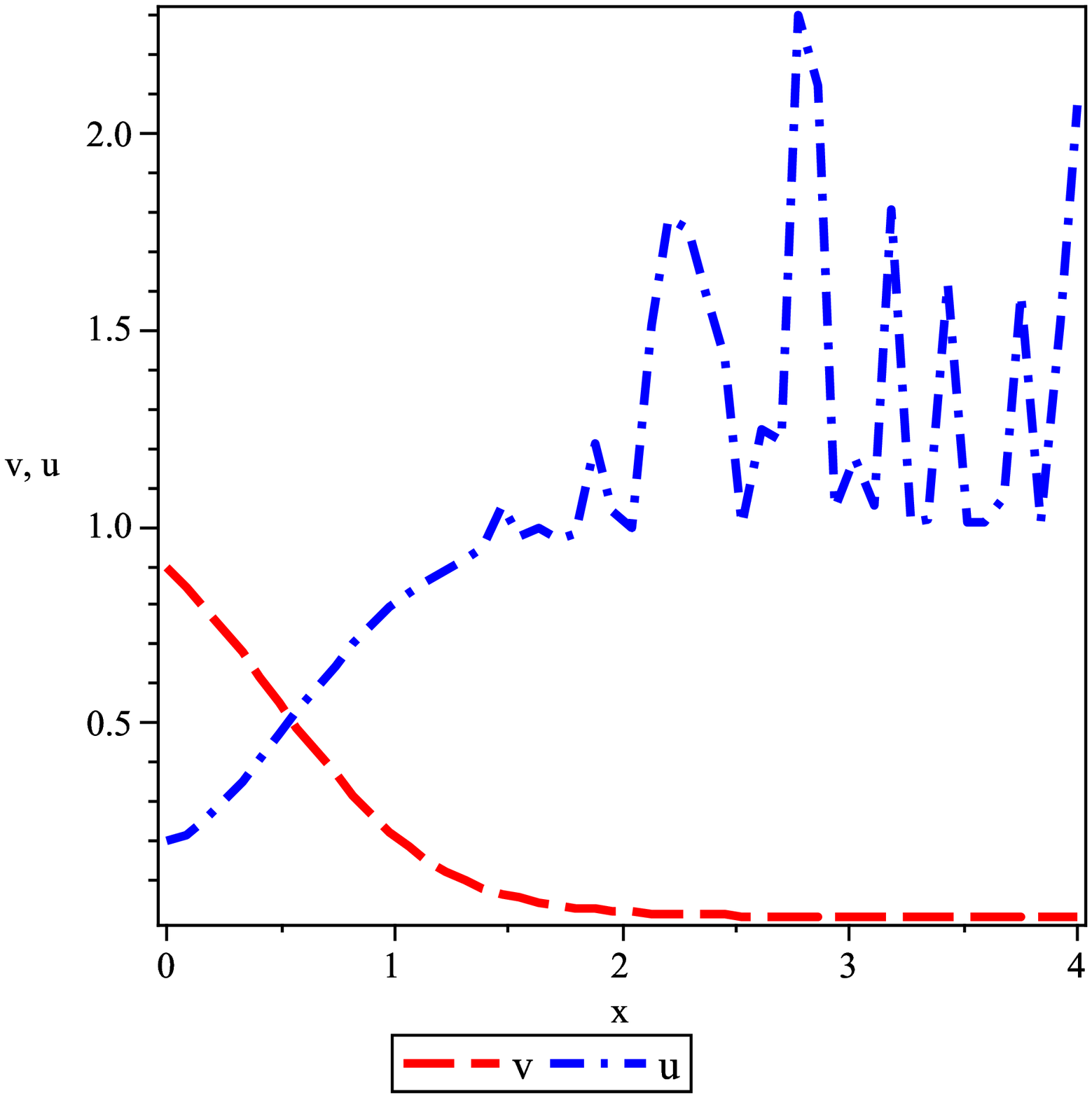}~~~~~\\
\vspace{1mm}
~~~~~~~~~~~~~~Fig. 1~~~~~~~~~~~~~~~~~~~~~~~~~~~~~~~~~~~~~~~~~Fig. 2~~~~~~~~~~~~~~~~~~~~~~~~~~~~~~~~~~~~~~~~Fig. 3~~~~~~~~~~~\\

\vspace{3mm}

\includegraphics[height=2in]{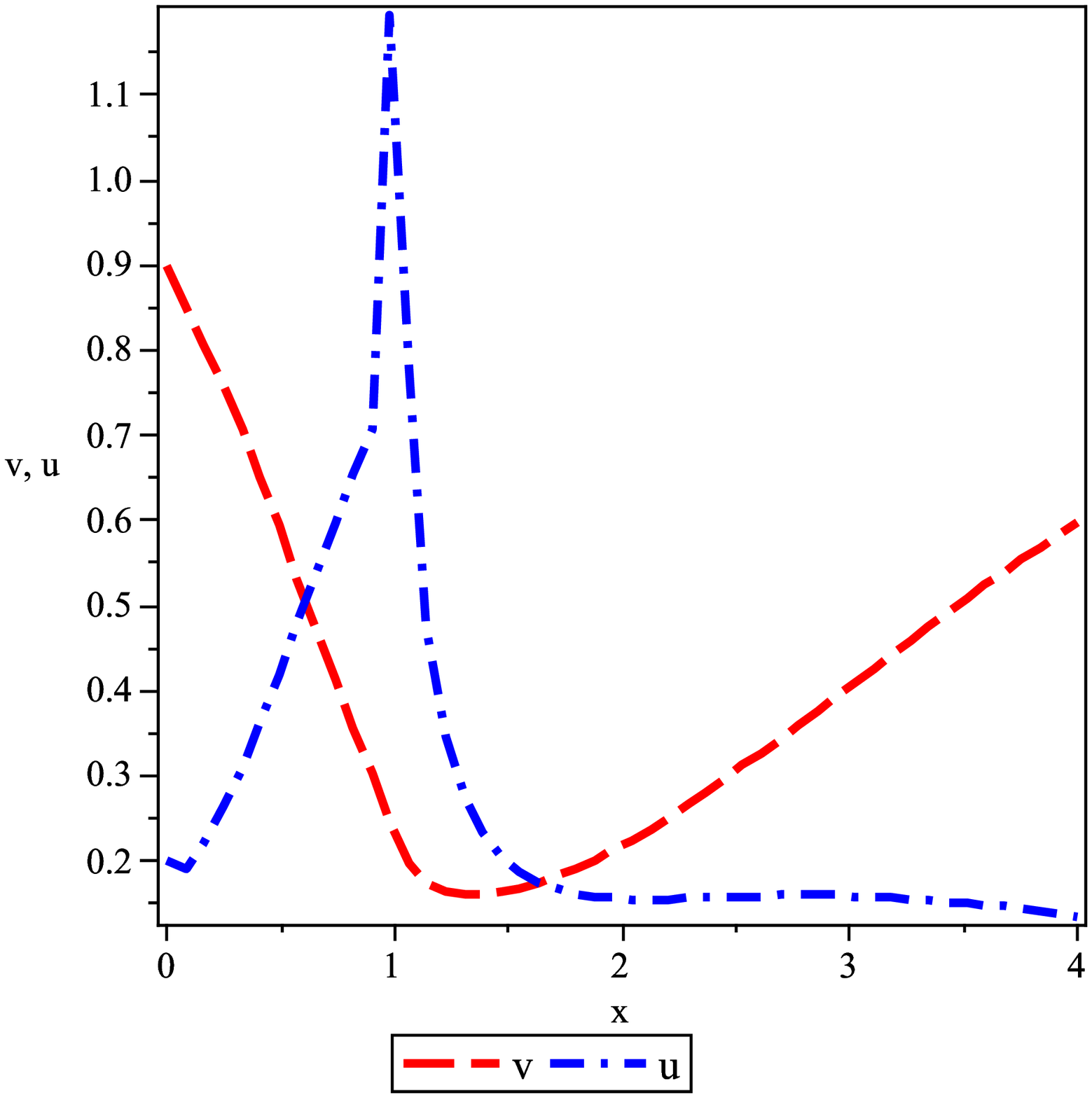}~~~~~~~~\includegraphics[height=2in]{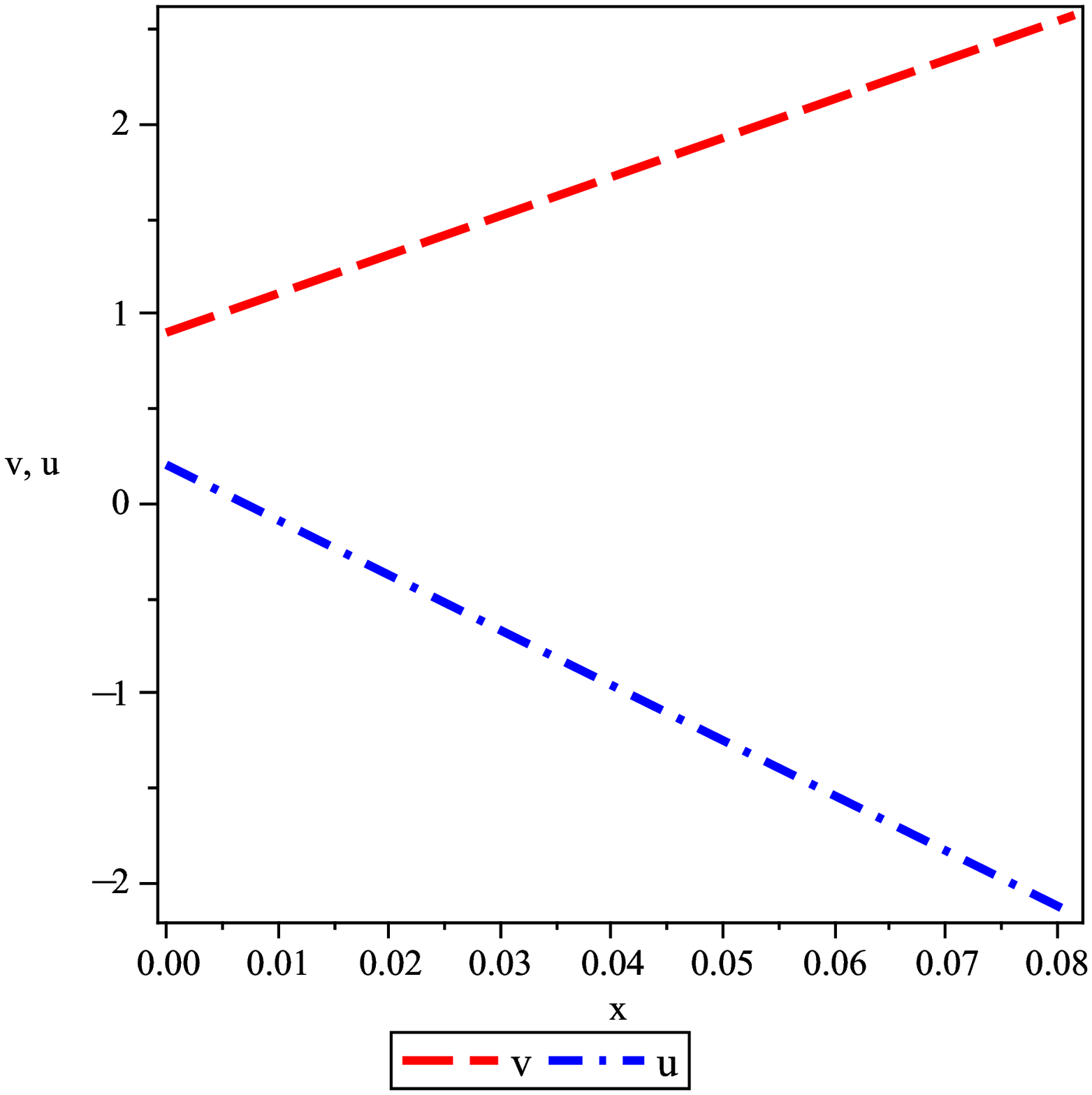}~~~~~~~~\includegraphics[height=2in]{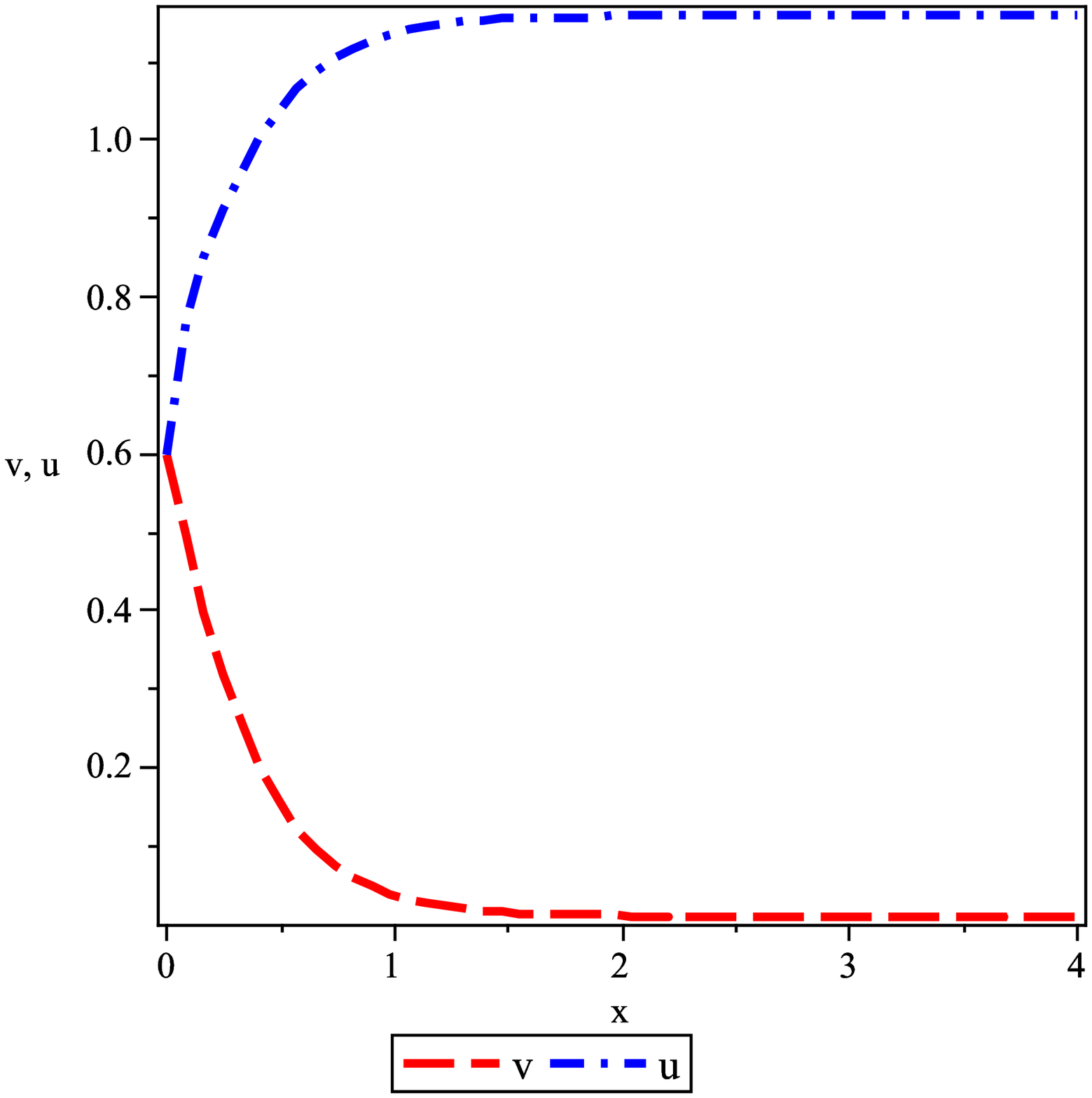}~~~~~~\\
\vspace{1mm}
~~~~~Fig. 4~~~~~~~~~~~~~~~~~~~~~~~~~~~~~~~~~~~~~~~~~~~~Fig. 5~~~~~~~~~~~~~~~~~~~~~~~~~~~~~~~~~~~~~~~~Fig. 6~~~~~~~~~\\

\vspace{3mm}

\includegraphics[height=2in]{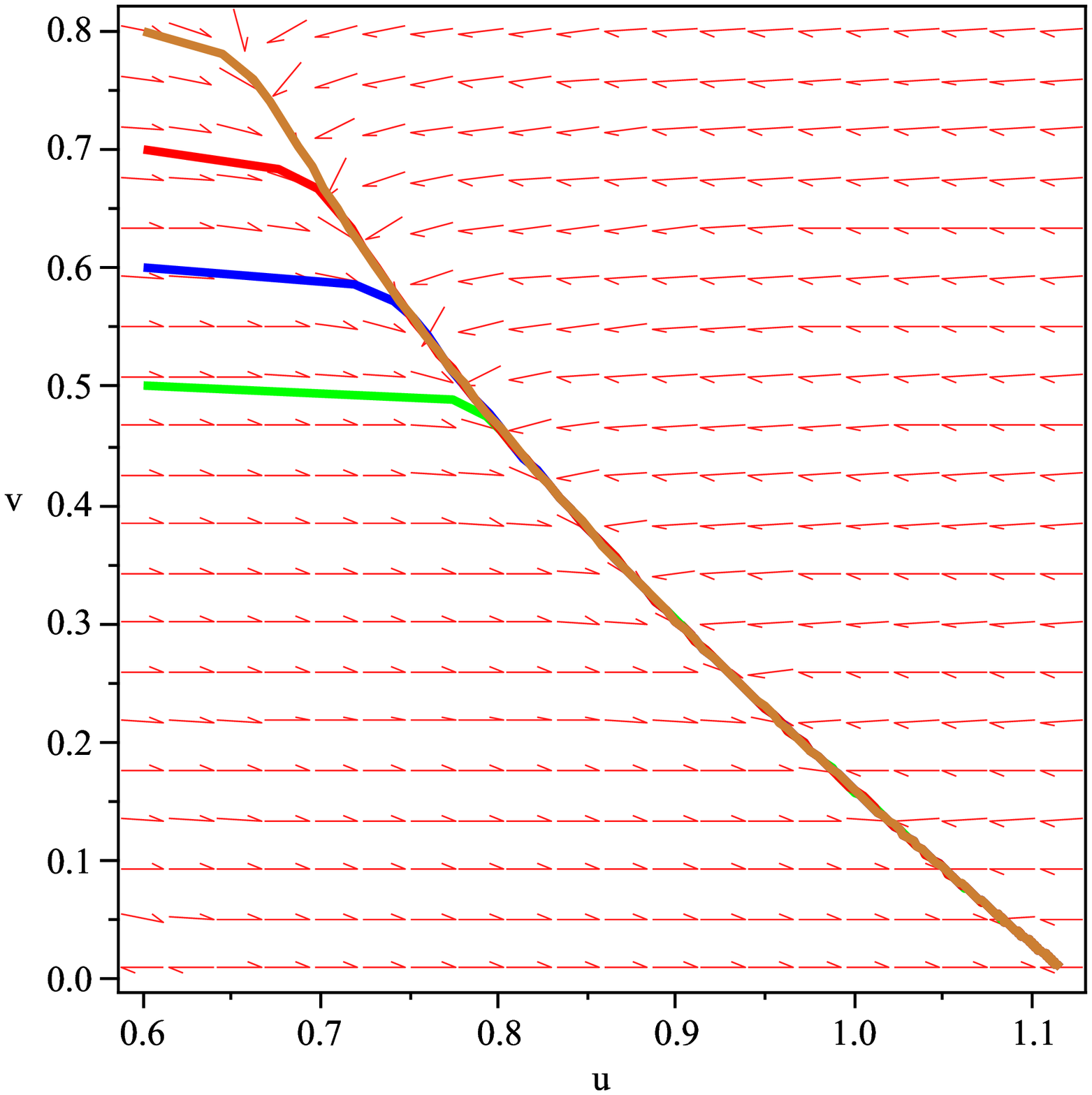}~~~~~~~~\includegraphics[height=2in]{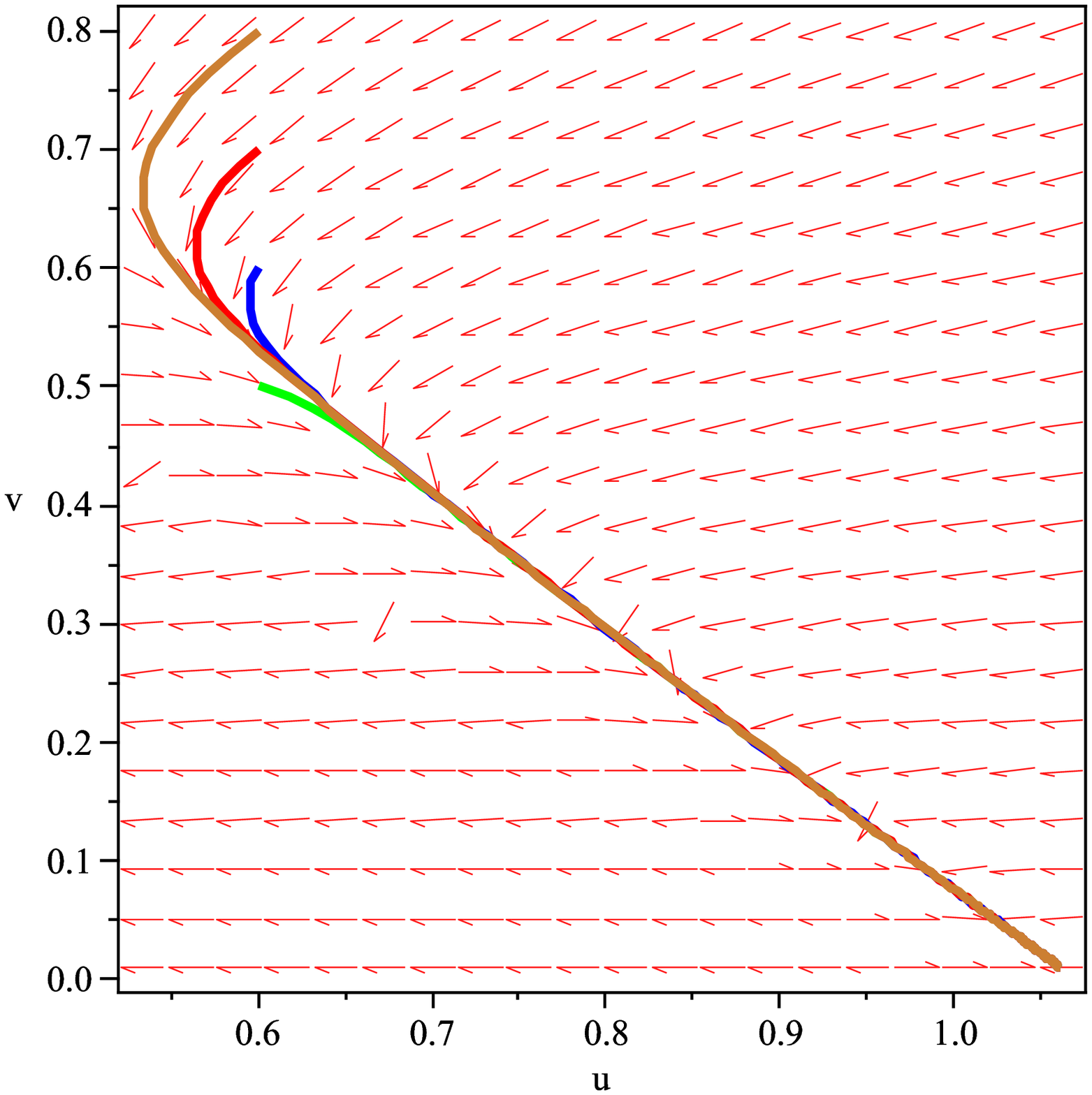}~~~~~~~~\includegraphics[height=2in]{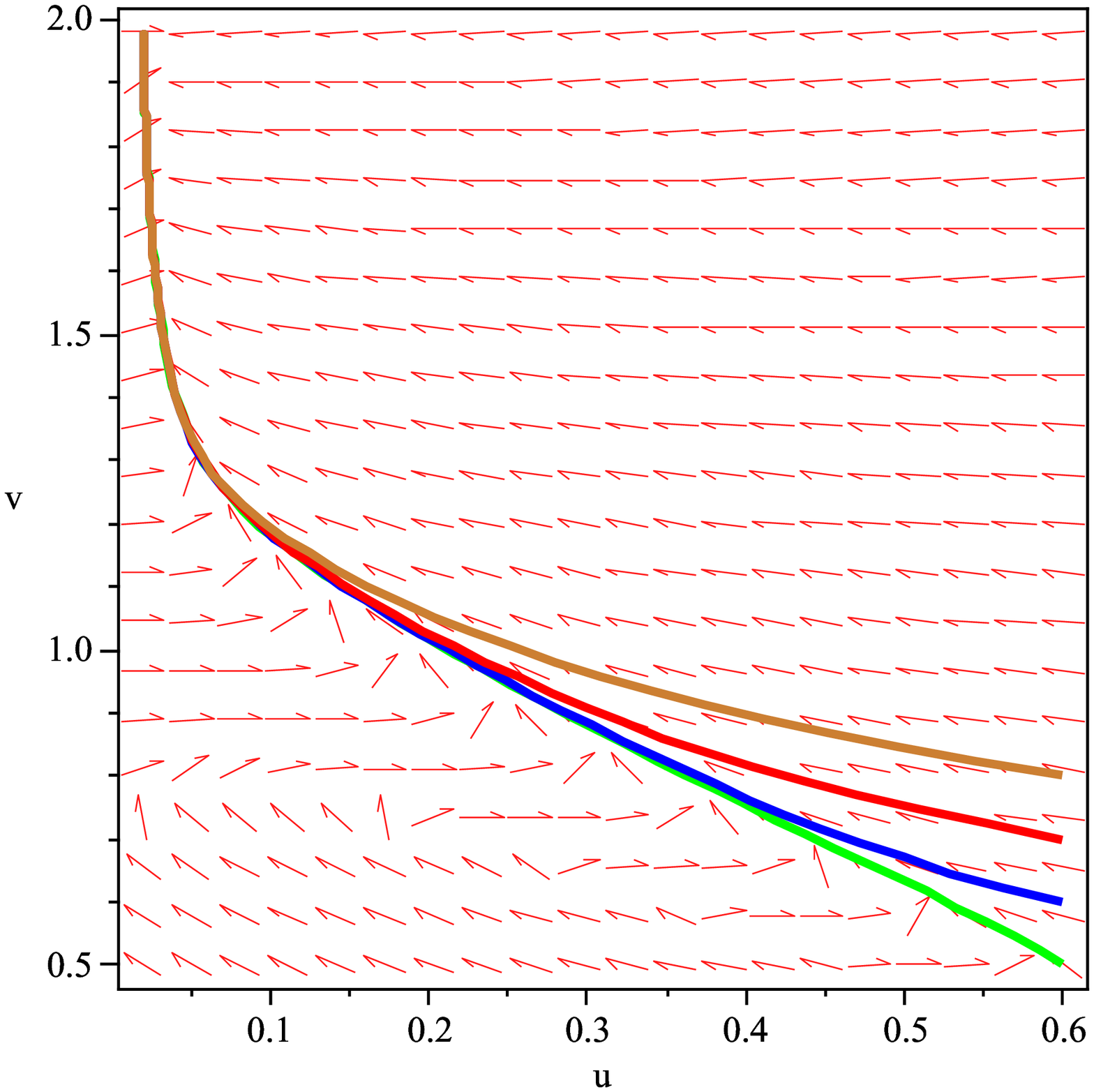}~~~~\\
\vspace{1mm}
~~~~Fig. 7~~~~~~~~~~~~~~~~~~~~~~~~~~~~~~~~~~~~~~~~~~~~~Fig. 8~~~~~~~~~~~~~~~~~~~~~~~~~~~~~~~~~~~~~~~~Fig. 9~~~~~~~\\

\vspace{3mm}

\includegraphics[height=1.5in]{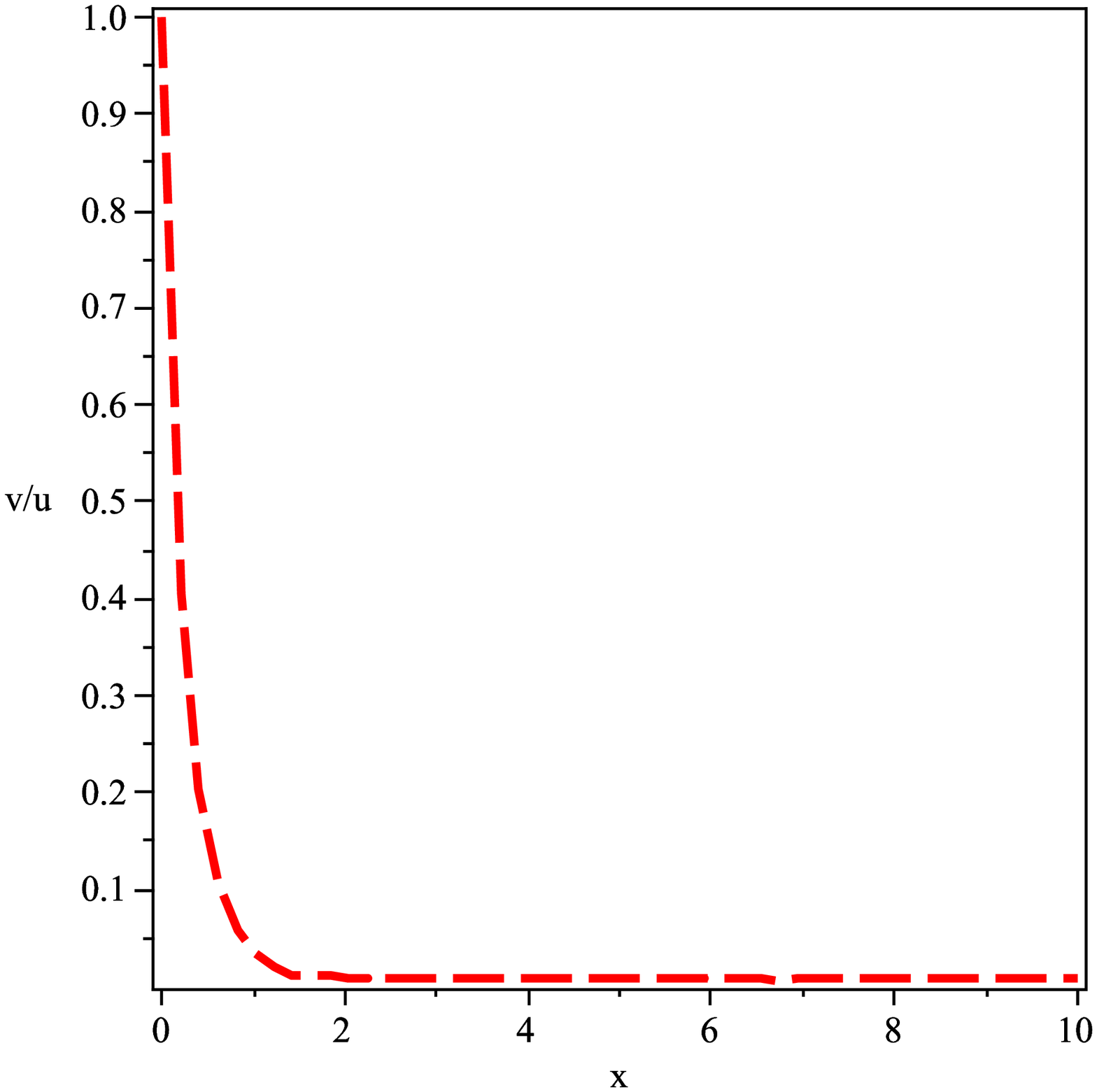}~~~~~~~~\includegraphics[height=1.5in]{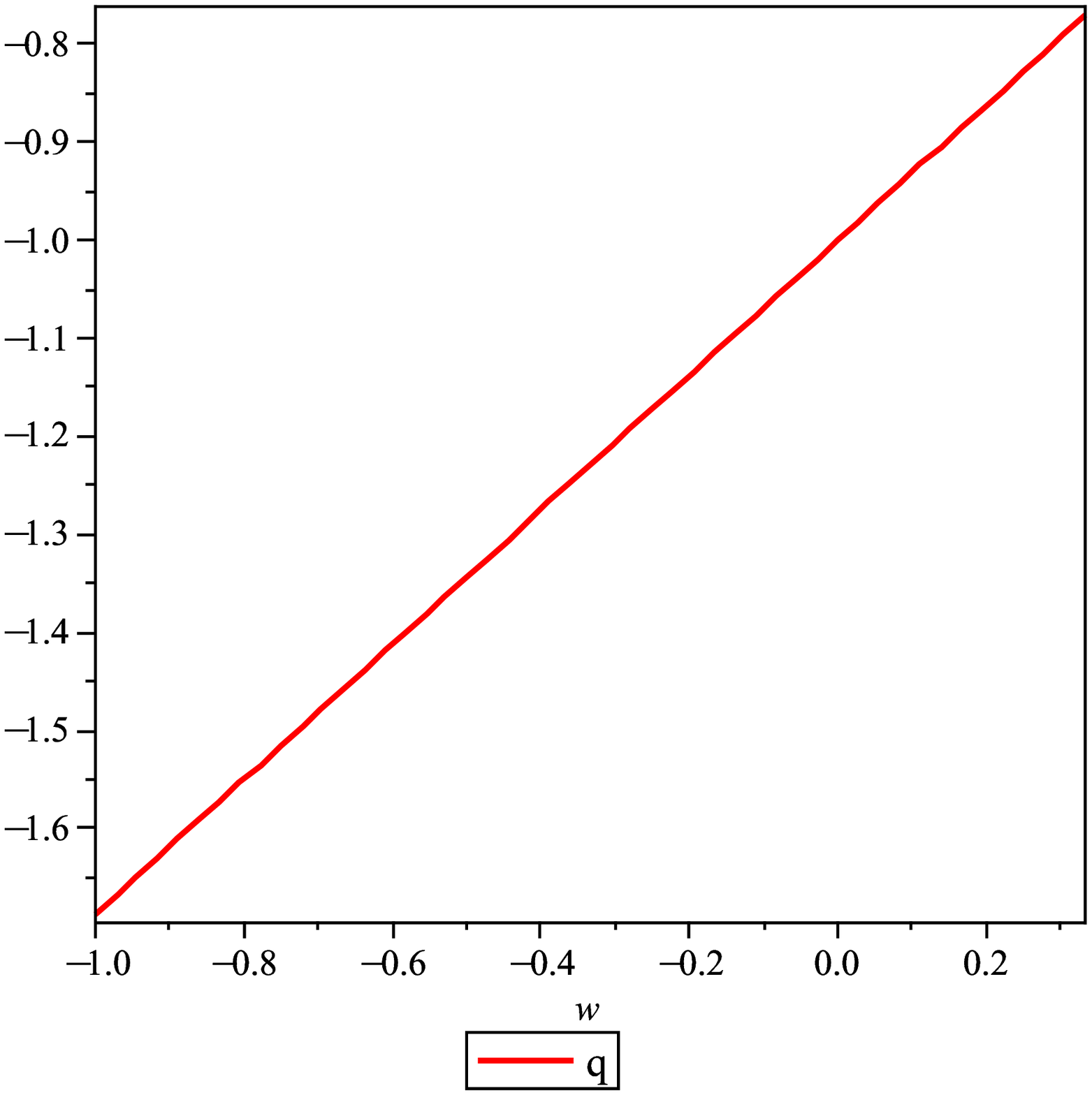}~~~~~~~~\includegraphics[height=1.5in]{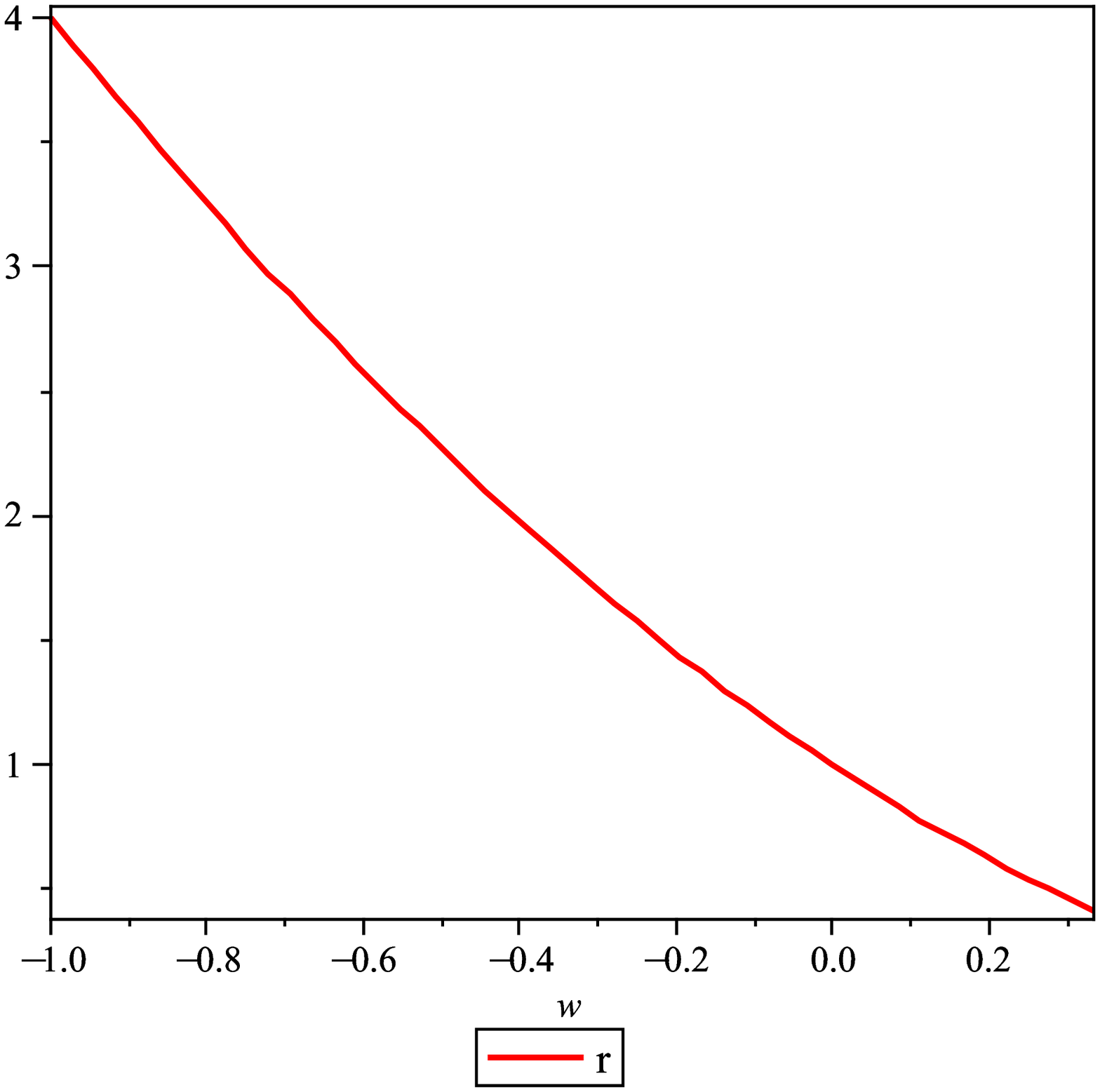}~~~~~~~\includegraphics[height=1.5in]{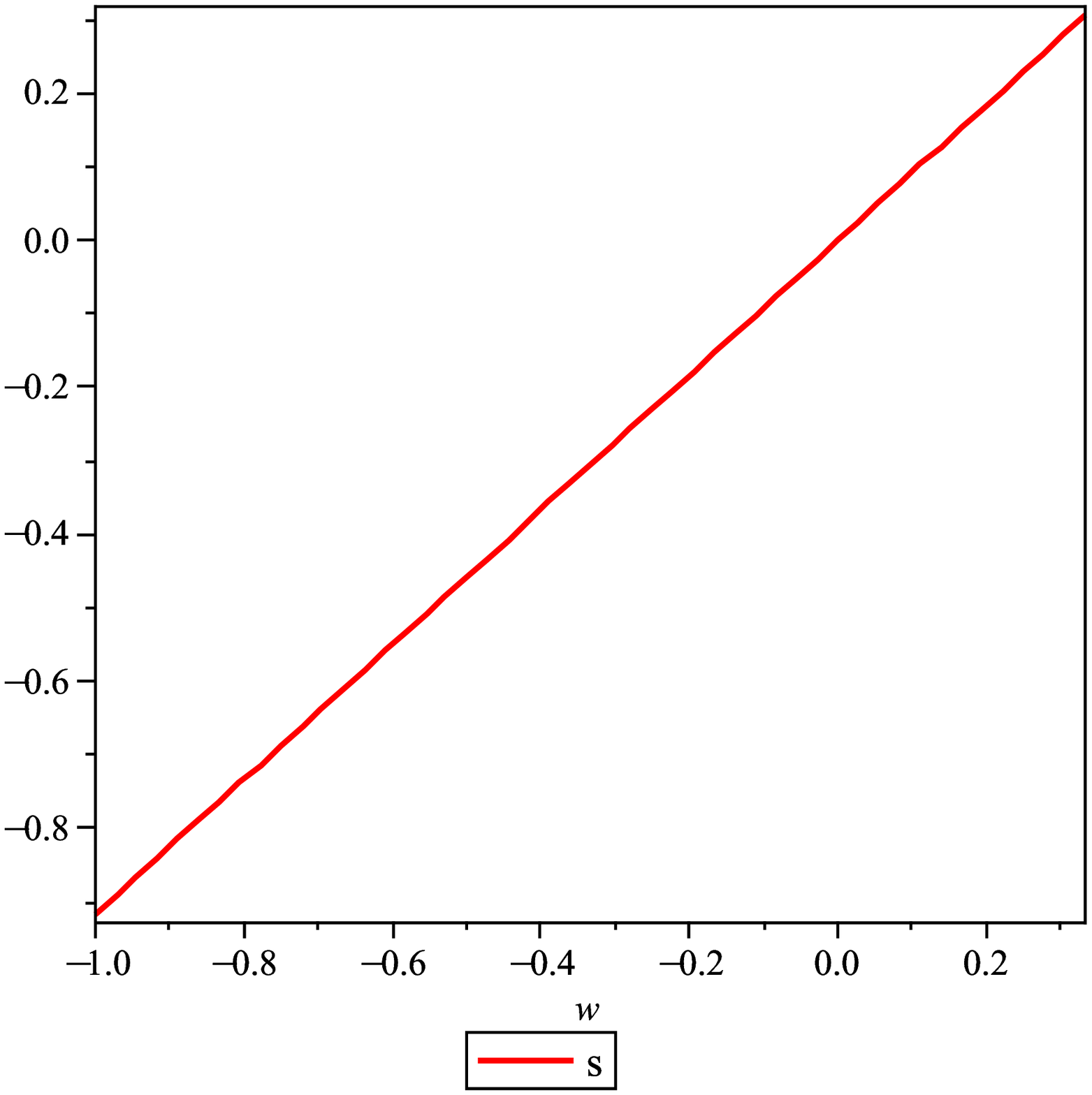}~\\
\vspace{1mm}
~~~~~~Fig. 10~~~~~~~~~~~~~~~~~~~~~~~~Fig. 11~~~~~~~~~~~~~~~~~~~~~~~~~~~~~Fig. 12~~~~~~~~~~~~~~~~~~~~~~~~~~Fig. 13\\

\end{figure}

\vspace{1mm}

{\bf Description of figures}

\vspace{2mm}

Figs.1, 2, 3, 4 : The dimensionless density parameters are plotted
against e-folding time for different values of brane tensions,
$\lambda$. The initial condition is $v(0)=0.9, u(0)=0.2$. The
other parameters are fixed at $w=-1, C=-1$, $\kappa_{4}=1$ and
$b=0.01$.
The brane tensions are respectively $\lambda=1, -10$, $-50$ and $-100$.\\

Fig.5 : The dimensionless density parameters are plotted against
e-folding time for a high value of interaction $(b=5)$. The
initial condition is $v(0)=0.9, u(0)=0.2$. The other
parameters are fixed at $w=-1, C=-1$, $\kappa_{4}=1$ and $\lambda=-10$.\\

Fig.6: The dimensionless density parameters are plotted against
e-folding time for same initial condition. The initial condition
is $v(0)=0.6, u(0)=0.6$. The other parameters are fixed at $w=-1,
C=-1$, $\kappa_{4}=1$ and $\lambda=-5$.\\

Figs.7, 8 : The phase diagram of the parameters depicting an
attractor solution are obtained for different values of brane
tension. The initial conditions chosen are $v(0)=0.5, u(0)=0.6$
(green); $v(0)=0.6, u(0)=0.6$ (blue); $v(0)=0.7, u(0)=0.6$ (red);
$v(0)=0.8, u(0)=0.6$ (brown). Other parameters are fixed at The
other parameters are fixed at $w=-1, C=-1$,
$\kappa_{4}=1$ and $b=0.01$. The brane tensions are respectively $\lambda=1, -10$. \\

Fig.9: The phase diagram of the parameters depicting an attractor
solution are obtained for a high value of interaction $(b=1)$. The
initial conditions chosen are $v(0)=0.5, u(0)=0.6$ (green);
$v(0)=0.6, u(0)=0.6$ (blue); $v(0)=0.7, u(0)=0.6$ (red);
$v(0)=0.8, u(0)=0.6$ (brown). Other parameters are fixed at The
other parameters are fixed at $w=-1, C=-1$,
$\kappa_{4}=1$ and $\lambda=-10$. \\

Fig. 10 : The ratio of density parameters is shown against
e-folding time. The initial conditions chosen are v(0)=0.6,
u(0)=0.6. The other parameters are fixed at The other parameters
are fixed at $w=-1, C=-1$, $\lambda=1$,
$\kappa_{4}=1$ and $b=0.01$.\\

Fig. 11 :The deceleration parameter is plotted against the EoS
parameter. Other parameters are fixed at $\psi_{(RSII)}=-0.5$.\\
Figs.12, 13 : The statefinder parameter $r$ and $s$ are plotted
against the EoS parameter. Other parameters are fixed at $\psi_{(RSII)}=-0.5$.\\

\subsubsection{\bf NATURE OF COSMOLOGICAL PARAMETERS}

\noindent

{\bf 1. Deceleration Parameter: }

\vspace{2mm}

The deceleration parameter, $q=-1-\frac{\dot{H}}{H^{2}}$ in this
model is calculated as,

\begin{equation}
q^{(RSII)}=-1+\frac{3}{2}\frac{\left\{\frac{\rho}{2\lambda}\left(\omega_{gccg}^{(RSII)}\frac{\rho_{gccg}}{\rho}+2\right)+\left(1+\omega_{gccg}^{(RSII)}\frac{\rho_{gccg}}{\rho}\right)\right\}}{\left(1+\frac{\rho}{2\lambda}\right)}
\end{equation}

We consider a dimensionless density parameter
$\Omega_{gccg}=\frac{\rho_{gccg}}{\rho}$. In terms of this density
parameter the expression for the deceleration parameter,
$q^{(RSII)}$ can be rewritten as,

\begin{equation}
q^{(RSII)}=-1+\frac{3}{2}\frac{\left\{\left(\frac{\rho}{2\lambda}+1\right)\omega_{gccg}^{(RSII)}\Omega_{gccg}+\left(1+\frac{\rho}{\lambda}\right)\right\}}{\left(1+\frac{\rho}{2\lambda}\right)}
\end{equation}

It is evident that for $\lambda\rightarrow \infty$, we retrieve
the result for the Einstein gravity,

\begin{equation}
q_{EG}=-1+\frac{3}{2}\left(1+\omega_{gccg}^{(RSII)}\Omega_{gccg}\right)
\end{equation}

Now since  $\Omega_{gccg}=\frac{\rho_{gccg}}{\rho}=\frac{u}{u+v}$
and assuming $\frac{\rho}{2\lambda}=\psi_{(RSII)}$ we get,

\begin{equation}
q^{(RSII)}=-1+\frac{3}{2}\frac{\left\{\left(1+\psi_{(RSII)}\right)\omega_{gccg}^{(RSII)}\frac{u}{u+v}
+\left(1+2\psi_{(RSII)}\right)\right\}}{\left(1+\psi_{(RSII)}\right)}
\end{equation}

Now at the critical point, $(u,v)\rightarrow(u_{c},v_{c})$, and
using equation (27) we get

\begin{equation}
q_c^{(RSII)}=-1+\frac{3}{2}Z_{(RSII)},~~~ where ~~~~
Z_{(RSII)}=\frac{\left\{\left(1+\psi_{(RSII)}\right)\omega_{gccg}^{(RSII)}\frac{u_{c}}{u_{c}+v_{c}}
+\left(1+2\psi_{(RSII)}\right)\right\}}{\left(1+\psi_{(RSII)}\right)}
\end{equation}

\noindent

{\bf Case I:} \vspace{1mm}

If~~
$$\psi_{(RS II)}=-\frac{\left(\omega_{gccg}+1\right)u+v}{\left(\omega_{gccg}+2\right)+2v}$$
then $Z_{(RS II)}=0$. So we have $q=-1$, which confirms the
accelerated expansion of the universe.\\

\noindent

{\bf Case II:} \vspace{1mm}

If
$$\psi_{(RS II)}=-1,~~~ $$
then $Z_{(RSII)}\rightarrow -\infty$. Therefore
$q\rightarrow-\infty$. Hence in this case we have the super
accelerated expansion of the universe.\\

\noindent

{\bf Note:}~In both the above cases we see that $\psi_{(RS
II)}<0$.~ We know that $\psi_{(RS II)}=\frac{\rho}{2\lambda}$.
Since the energy density, $\rho$ is always positive, therefore
$\lambda$ should be negative ($\lambda<0$). {\bf Hence in order to
realize the recent cosmic acceleration the  RS II brane model
should possess a negative brane tension}. It is known from
\cite{Maartens1} standard model fields are confined on the
negative tension (or "visible") brane rather than the positive
tension ("hidden") brane of the RS II model. Hence our result is
consistent with the basic idea of formulation of the RS II brane
model.

\noindent

\vspace{1mm}

In this scenario, the Hubble parameter can be obtained as,
\begin{equation}
H=\frac{2}{3Z_{(RSII)}t}
\end{equation}
where the integration constant has been ignored. Integration of
equation (29) yields
\begin{equation}
a(t)=a_{0}t^{\frac{2}{3Z_{(RSII)}}}
\end{equation}
which gives the power law form of expansion of the universe. In
order to have an accelerated expansion of universe we must have
$0<Z_{(RSII)}<\frac{2}{3}$. Using this range of $Z_{(RSII)}$ in
the equation $q_{c}^{RSII}=-1+\frac{3}{2}Z_{(RSII)}$, i.e., eqn.
(28), we get the range of $q_{c}^{(RSII)}$ as
$-1<q_{c}^{(RSII)}<0$. This is again consistent with an
accelerated expansion of the universe.

\vspace{2mm}

{2. \bf Statefinder Parameters }

\vspace{2mm}

\noindent

As so many cosmological models have been developed, so for
discrimination between these contenders, Sahni et al \cite{Sahni1}
proposed a new geometrical diagnostic named the statefinder pair
$\left\{r, s\right\}$, where $r$ is generated from the scale
factor $a$ and its derivatives with respect to the cosmic time $t$
upto the third order and $s$ is a simple combination of $r$ and
the deceleration parameter $q$. Clear differences for the
evolutionary trajectories in the $r-s$ plane have been found
\cite{j4,j5,j6}. The statefinder parameters are defined as
follows,
\begin{equation}
r\equiv\frac{\stackrel{...}a}{aH^3},\ \
~~~~~~~~s\equiv\frac{r-1}{3(q-1/2)}.
\end{equation}
The expressions for the statefinder pair (eqn.(31)) in the RS II
model can be obtained in the form
\begin{equation}
r_{(RSII)}=\left(1-\frac{3Z_{(RSII)}}{2}\right)\left(1-3Z_{(RSII)}\right).
\end{equation}
and
\begin{equation}
s_{(RSII)}=Z_{(RSII)}
\end{equation}

\noindent

The trajectories in the $r - s$ plane for various existing models
can exhibit quite different behaviours. The deviation of these
trajectories from the $(0, 1)$ point defines the distance of a
given model from the $\Lambda$CDM model. The statefinder pair
$\{r, s\}$ can successfully differentiate between a wide variety
of cosmological models including a cosmological constant,
quintessence, Chaplygin gas, and interacting dark energy models.
In a given model the pair $\{r, s\}$ can be computed and the
trajectory in the $r - s$ plane can be drawn. Furthermore, the
values of $r$, $s$ can be extracted from future observations
\cite{Albert1, Albert2}.

{\bf Note:~ It is quite interesting to note that the pair
$\{r_{(RSII)},~ s_{(RSII)}\}$ yields the $\Lambda$CDM model
$\{r_{EG},~ s_{EG}\}=\{1,0\}$ when $Z_{(RSII)}=0$}.

\section{MODEL 2: DGP BRANE MODEL}

\noindent

A simple and effective model of brane-gravity is the
Dvali-Gabadadze-Porrati (DGP) braneworld model \cite{dgp1, Def1,
Def2} which models our 4-dimensional world as a FRW brane embedded
in a 5-dimensional Minkowski bulk. It explains the origin of DE as
the gravity on the brane leaking to the bulk at large scale. On
the 4-dimensional brane the action of gravity is proportional to
$M_p^2$ whereas in the bulk it is proportional to the
corresponding quantity in 5-dimensions. The model is then
characterized by a cross over length scale $
r_c=\frac{M_p^2}{2M_5^2} $ such that gravity is 4-dimensional
theory at scales $a<<r_c$ where matter behaves as pressureless
dust, but gravity leaks out into the bulk at scales $a>>r_c$ and
matter approaches the behaviour of a cosmological constant.
Moreover it has been shown that the standard Friedmann cosmology
can be firmly embedded in DGP brane.

It may be noted that in literature, standard DGP model has been
generalized to (i) LDGP model by adding a cosmological constant
\cite{Lue1}, (ii) QDGP model by adding a quintessence perfect
fluid \cite{Chimento1}, (iii) CDGP model by Chaplygin gas
\cite{Bou1} and (iv) SDGP by a scalar field \cite{Zhang2}. In
\cite{Wu2} the DGP model has been analysed by adding Holographic
DE (HDE).

\noindent

While flat, homogeneous and isotropic brane is being considered,
the Friedmann equation in DGP brane model \cite{dgp1, Def1, Def2}
is modified to the equation
\begin{equation}
H^2=\left(\sqrt{\frac{\rho}{3}+\frac{1}{4r_{c}^{2}}}+\epsilon
\frac{1}{2r_c}\right)^2,
\end{equation}
where $H=\frac{\dot a}{a}$ is the Hubble parameter, $\rho$ is the
total cosmic fluid energy density and $r_c=\frac{M_p^2}{2M_5^2}$
is the cross-over scale which determines the transition from 4D to
5D behaviour and $\epsilon=\pm 1 $ (choosing $M_{p}^{2}=8\pi
G=1$). For $\epsilon=+1$, we have standard DGP$(+)$ model which is
self accelerating model without any form of DE, and effective $w$
is always non-phantom. However for $\epsilon=-1$, we have DGP$(-)$
model which does not self accelerate but requires DE on the brane.
It experiences 5D gravitational modifications to its dynamics
which effectively screen DE. Brane world scenario is actually a
modified gravity theory. If we write the Einstein equation for
brane world in terms of Einstein gravity then the extra term can
be treated as the effective DE. But that is not the physical DE.
Moreover this DE is applicable only in Einstein gravity. But here
we will consider the physical DE in brane world. So we have
introduced the GCCG type fluid in brane.

Consequently using the Friedmann equation (34) and the
conservation equations, we obtain the modified Raychaudhuri
equation
\begin{equation}
\left(2H-\frac{\epsilon}{r_c}\right)\dot{H}=-H\left(\rho+p\right),
\end{equation}

\subsection{DYNAMICAL SYSTEM ANALYSIS}

\noindent

Just like the previous model, here also we proceed to perform the
dynamical system analysis for DGP brane model:\\

The system is obtained by using the equations (1), (9), (10),
(11), (34) and(35) as given below,

\begin{equation}
\frac{du}{dx}=-3b\left(u+v\right)-3u\left(1+\omega_{gccg}^{DGP}\right)+\frac{2r_{c}^{4}\left(1-u-v\right)^{4}}{3\left(1+u+v\right)}\left[\frac{9u\left(u+v\right)}{r_{c}^{4}\left(1-u-v\right)^{4}}-\left\{C+\left(\frac{9u^{2}}{r_{c}^{4}\left(1-u-v\right)^{4}}-C\right)^{-w}\right\}\right]
\end{equation}

\begin{equation}
\frac{dv}{dx}=3b\left(u+v\right)-3v+\frac{2r_{c}^{4}v\left(1-u-v\right)^{4}}{3u\left(1+u+v\right)}\left[\frac{9u\left(u+v\right)}{r_{c}^{4}\left(1-u-v\right)^{4}}-\left\{C+\left(\frac{9u^{2}}{r_{c}^{4}\left(1-u-v\right)^{4}}-C\right)^{-w}\right\}\right]
\end{equation}

Where, $\omega_{gccg}^{DGP}$ is the EoS parameter for GCCG in DGP
brane determined as,
\begin{equation}
\omega_{gccg}^{DGP}=\frac{p_{gccg}}{\rho_{gccg}}=-\frac{r_{c}^{4}\left(1-u-v\right)^{4}}{9u^{2}}\left[C+\left\{\frac{9u^{2}}{r_{c}^{4}\left(1-u-v\right)^{4}}-C\right\}^{-w}\right]
\end{equation}

\vspace{1mm}

For mathematical simplicity, here also we have considered
$\alpha=1$.\\

{\bf Note: ~It is to be noted that $\epsilon$ has been considered
as $-1$ (DGP(-) model). Hence it does not produce a
self-accelerating scenario of the brane model, but needs GCCG  as
dark energy (interacting) in order to realize the accelerating
scenario. Thus the introduction of interacting GCCG in DGP brane
is properly justified.}

\subsubsection{\bf CRITICAL POINTS}

\noindent

The critical points for the above system (eqns. (36) and (37)) are
calculated by putting $\frac{du}{dx}=0=\frac{dv}{dx}$. Here also
due to highly complicated forms of the equations, it is difficult
to get an explicit solution in terms of all the parameters. So we
find the following solution in terms of the interaction parameter
as below.

\begin{equation}
u_{1c}=\frac{1}{2}\left(1+\sqrt{1-4b}\right),~~~~~~~~~~~~~~~~v_{1c}=\frac{1}{2}\left(1-\sqrt{1-4b}\right)
\end{equation}

\begin{equation}
u_{2c}=\frac{1}{2}\left(1-\sqrt{1-4b}\right),~~~~~~~~~~~~~~~~v_{2c}=\frac{1}{2}\left(1+\sqrt{1-4b}\right)
\end{equation}

The other variables are taken as:

$$w=-1,~~~~~~ C=-1,~~~~~~ r_{c}=1,~~~~~~$$
It is obvious from the above values that the critical point exists
only for $b\leq\frac{1}{4}$. The critical point correspond to the
era dominated by DM and GCCG type DE. For the critical point
$(u_{ic},v_{ic}), ~~i=1,~2$, the equation of state parameter (eqn.
38) of the interacting DE takes the form:

\begin{equation}
\omega_{gccg}^{DGP}=-\frac{r_{c}^{4}\left(1-u_{ic}-v_{ic}\right)^{4}}{9u_{ic}^{2}}\left[C+\left\{\frac{9u_{ic}^{2}}{r_{c}^{4}\left(1-u_{ic}-v_{ic}\right)^{4}}-C\right\}^{-w}\right]
\end{equation}
where $i=1, ~2$

\subsubsection{\bf STABILITY AROUND CRITICAL POINT}

\noindent

Now if we write $\widehat{f}=\frac{du}{dx}$ and
$\widehat{g}=\frac{dv}{dx}$, then we can obtain the following
expressions
\begin{equation}
\delta\left(\frac{du}{dx}\right)=\left[\partial_{u}
\widehat{f}\right]_{c}\delta u+\left[\partial_{v}
\widehat{f}\right]_{c}\delta v
\end{equation}
and
\begin{equation}
\delta\left(\frac{dv}{dx}\right)=\left[\partial_{u}
\widehat{g}\right]_{c}\delta u+\left[\partial_{v}
\widehat{g}\right]_{c}\delta v
\end{equation}
where

\vspace{2mm}

$$\partial_{u}\widehat{f}=\frac{1}{3u^{2}\left(u+v+1\right)^{2}\left\{-9u^{2}+Cr_{c}^{4}\left(u+v-1\right)^{4}\right\}}\left\{-C+\frac{9u^{2}}{r_{c}^{4}\left(u+v-1\right)^{4}}\right\}^{-w}\left[-C^{2}r_{c}^{8}\left(u+v-1\right)^{7}\right.$$
$$\left.\left\{3u^{3}+\left(v-1\right)\left(v+1\right)^{2}+u^{2}\left(v+3\right)-u\left(v+1\right)\left(v+5\right)\right\}\left\{-C++\frac{9u^{2}}{r_{c}^{4}\left(u+v-1\right)^{4}}\right\}^{w}-Cr_{c}^{4}\left(u+v-1\right)^{3}\right.$$
$$\left.\left\{r_{c}^{4}\left(u+v-1\right)^{4}\left(3u^{3}+\left(v-1\right)\left(v+1\right)^{2}+u^{2}\left(v+3\right)-u\left(v+1\right)\left(v+5\right)\right)+9u^{2}\left(-C+\frac{9u^{2}}{r_{c}^{4}\left(u+v-1\right)^{4}}\right)^{w}\right.\right.$$
$$\left.\left.\left(b\left(u+v-1\right)\left(u+v+1\right)^{2}-2\left(2u^{3}-v+v^{3}+2u^{2}\left(v+1\right)+u\left(v-4\right)\left(v+1\right)\right)\right)\right\}+9u^{2}\left\{9u^{2}\left(-C+\right.\right.\right.$$
$$\left.\left.\left.\frac{9u^{2}}{r_{c}^{4}\left(u+v-1\right)^{4}}\right)^{w}\left(1-2u-\left(u+v\right)^{2}+b\left(u+v+1\right)^{2}\right)+r_{c}^{4}\left(u+v-1\right)^{3}\left(\left(v-1\right)\left(v+1\right)^{2}\left(1+2w\right)\right.\right.\right.$$
\begin{equation}
\left.\left.\left.+u^{3}\left(3+2w\right)+u^{2}\left(3+v+2w-2vw\right)-u\left(v+1\right)\left(5+v+2\left(v+1\right)w\right)\right)\right\}\right]
\end{equation}

\vspace{2mm}

$$\partial_{v}\widehat{f}=\frac{1}{3}\left[-9b-\frac{2\left\{-C+\frac{9u^{2}}{r_{c}^{4}\left(u+v-1\right)^{4}}\right\}^{-w}}{u\left(u+v+1\right)^{2}\left\{-9u^{2}+Cr_{c}^{4}\left(u+v-1\right)^{4}\right\}}\left\{C^{2}r_{c}^{8}\left(u+v-1\right)^{7}\left(u+u^{2}-uv-2\left(1+v\right)^{2}\right)\right\}\right.$$
$$\left.\left\{-C+\frac{9u^{2}}{r_{c}^{4}\left(u+v-1\right)^{4}}\right\}^{w}+Cr_{c}^{4}\left(u+v-1\right)^{3}\left\{r_{c}^{4}\left(u+v-1\right)^{4}\left(u+u^{2}-uv-2\left(v+1\right)^{2}\right)-9u^{2}\left(-3+2u\right.\right.\right.$$
$$\left.\left.\left.+u^{2}-\left(u+3\right)v-2v^{2}\right)\left(-C+\frac{9u^{2}}{r_{c}^{4}\left(u+v-1\right)^{4}}\right)^{w}\right\}-9u^{2}\left\{-9u^{2}\left(-C+\frac{9u^{2}}{r_{c}^{4}\left(u+v-1\right)^{4}}\right)^{w}\right.\right.$$
\begin{equation}
\left.\left.+r_{c}^{4}\left(u+v-1\right)^{3}\left(u-uv-2\left(v+1\right)^{2}\left(w+1\right)+u^{2}\left(1+2w\right)\right)\right\}\right]
\end{equation}

\vspace{2mm}

$$\partial_{u}\widehat{g}=\frac{1}{3u^{2}\left(u+v+1\right)^{2}\left\{-9u^{2}+Cr_{c}^{4}\left(u+v-1\right)^{4}\right\}}\left\{-C+\frac{9u^{2}}{r_{c}^{4}\left(u+v-1\right)^{4}}\right\}^{-w}\left[-2C^{2}r_{c}^{8}v\left(u+v-1\right)^{7}\right.$$
$$\left.\left(1+2u^{2}-v^{2}+u\left(v+5\right)\right)\left(-C+\frac{9u^{2}}{r_{c}^{4}\left(u+v-1\right)^{4}}\right)^{w}+Cr_{c}^{4}\left(u+v-1\right)^{3}\left(-2r_{c}^{4}v\left(u+v-1\right)^{4}\left(1+2u^{2}\right.\right.\right.$$
$$\left.\left.\left.-v^{2}+u\left(v+5\right)\right)+9u^{2}\left(-C+\frac{9u^{2}}{r_{c}^{4}\left(u+v-1\right)^{4}}\right)^{w}\left(b\left(u+v-1\right)\left(u+v+1\right)^{2}+2v\left(2u\left(u+3\right)+v+uv\right.\right.\right.\right.$$
$$\left.\left.\left.\left.-v^{2}\right)\right)\right)+9u^{2}\left\{-9u^{2}\left(-C+\frac{9u^{2}}{r_{c}^{4}\left(u+v-1\right)^{4}}\right)^{w}\left(2v+b\left(u+v+1\right)^{2}\right)+2r_{c}^{4}v\left(u+v-1\right)^{3}\left(2u^{2}\left(w+1\right)\right.\right.\right.$$
\begin{equation}
\left.\left.\left.-\left(v^{2}-1\right)\left(2w+1\right)+u\left(5+v+4w\right)\right)\right\}\right]
\end{equation}

\vspace{2mm}

$$\partial_{v}\widehat{g}=\frac{1}{3}\left[-9+9b-\frac{2r_{c}^{4}v\left(u+v-1\right)^{4}\left\{-C+\frac{9u\left(u+v\right)}{r_{c}^{4}\left(u+v-1\right)^{4}}-\left(-C+\frac{9u^{2}}{r_{c}^{4}\left(u+v-1\right)^{4}}\right)^{-w}\right\}}{u\left(u+v+1\right)^{2}}\right.$$
$$\left.+\frac{8r_{c}^{4}v\left(u+v-1\right)^{3}\left\{-C+\frac{9u\left(u+v\right)}{r_{c}^{4}\left(u+v-1\right)^{4}}-\left(-C+\frac{9u^{2}}{r_{c}^{4}\left(u+v-1\right)^{4}}\right)^{-w}\right\}}{u\left(u+v+1\right)}\right.$$
$$\left.+\frac{2r_{c}^{4}\left(u+v-1\right)^{4}\left\{-C+\frac{9u\left(u+v\right)}{r_{c}^{4}\left(u+v-1\right)^{4}}-\left(-C+\frac{9u^{2}}{r_{c}^{4}\left(u+v-1\right)^{4}}\right)^{-w}\right\}}{u\left(u+v+1\right)}\right.$$
\begin{equation}
\left.-\frac{18v\left\{1+3v+u\left(3+4w\left(-C+\frac{9u^{2}}{r_{c}^{4}\left(u+v-1\right)^{4}}\right)^{-1-w}\right)\right\}}{\left(u+v-1\right)\left(u+v+1\right)}\right]
\end{equation}

\noindent

The Jacobian matrix of the above system is given by,
$$
J_{\left(u,v\right)}^{(DGP)}=\left(\begin{array}{c}\frac{\delta
\widehat{f}}{\delta u} ~~~~~ \frac{\delta\widehat{f}}{\delta v}\\
\frac{\delta \widehat{g}}{\delta u}~~~~~ \frac{\delta
\widehat{g}}{\delta v}
\end{array}\right)
$$
Here we notice a very interesting feature of the model. We see
that $u_{c}+v_{c}=1$ for this model. Since the denominators of the
above partial derivatives contain the term $u+v-1$, hence they
become indeterminate at the critical point, leading to a highly
unstable scenario (chaos). As a result, determination of eigen
values at the critical point is not possible for this model. So we
resort to an alternative technique for our evaluations. We will
consider a very small neighbourhood of the critical point, thus
avoiding the chaos, and then try to calculate the eigen values at
any convenient point in the neighbourhood sufficiently close to
the critical point. This is purely based on the assumption that a
sufficiently close neighbouring point will retain most of the
properties of the critical point except the indeterminate nature.
Our evaluations led us to the following eigen values for the given
system:\\
${\bf \lambda_{1}=-2.97091,~~~ \lambda_{2}=0.734899}$. Hence it is
a saddle point.

\begin{figure}

\includegraphics[height=2in]{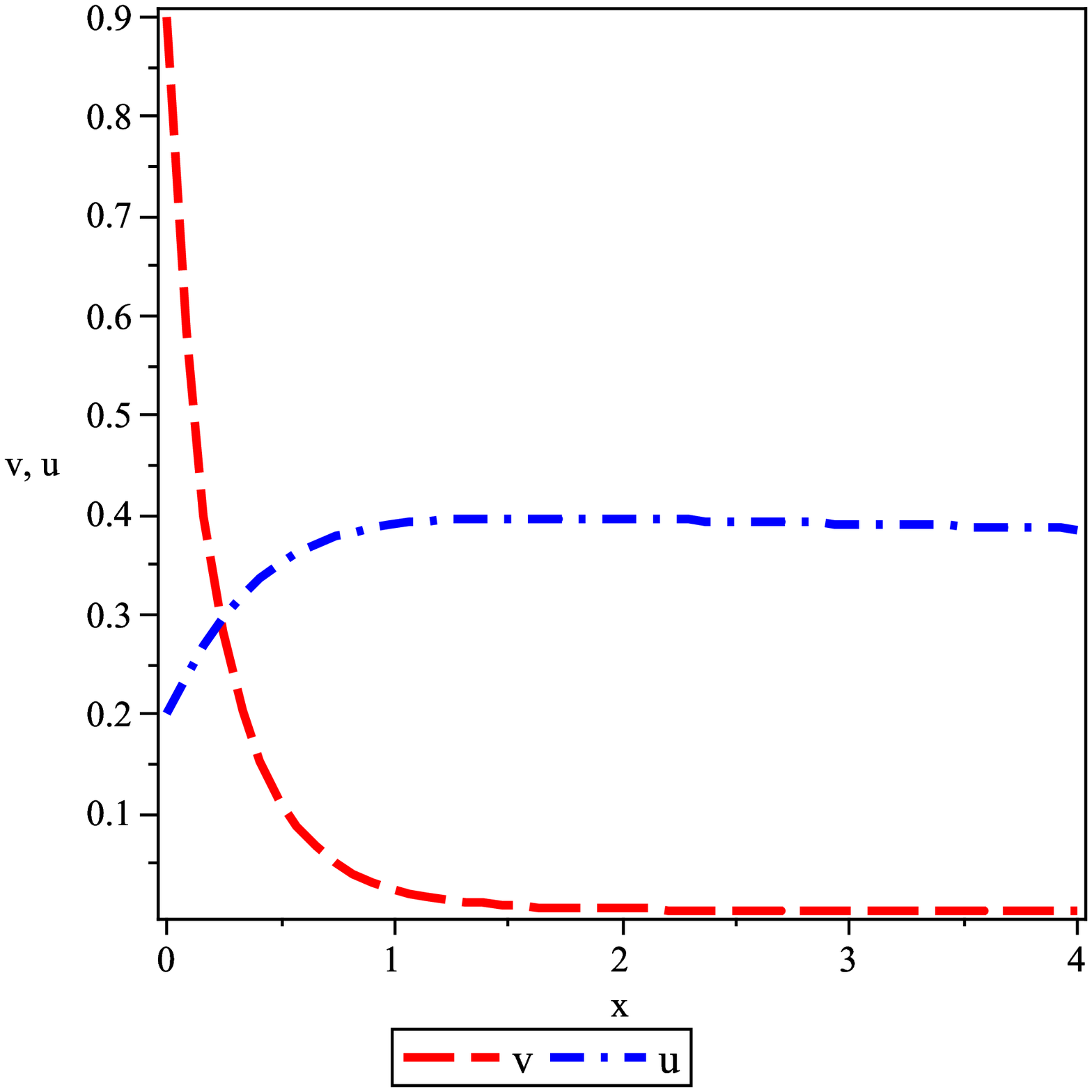}~~~~~~~~\includegraphics[height=2in]{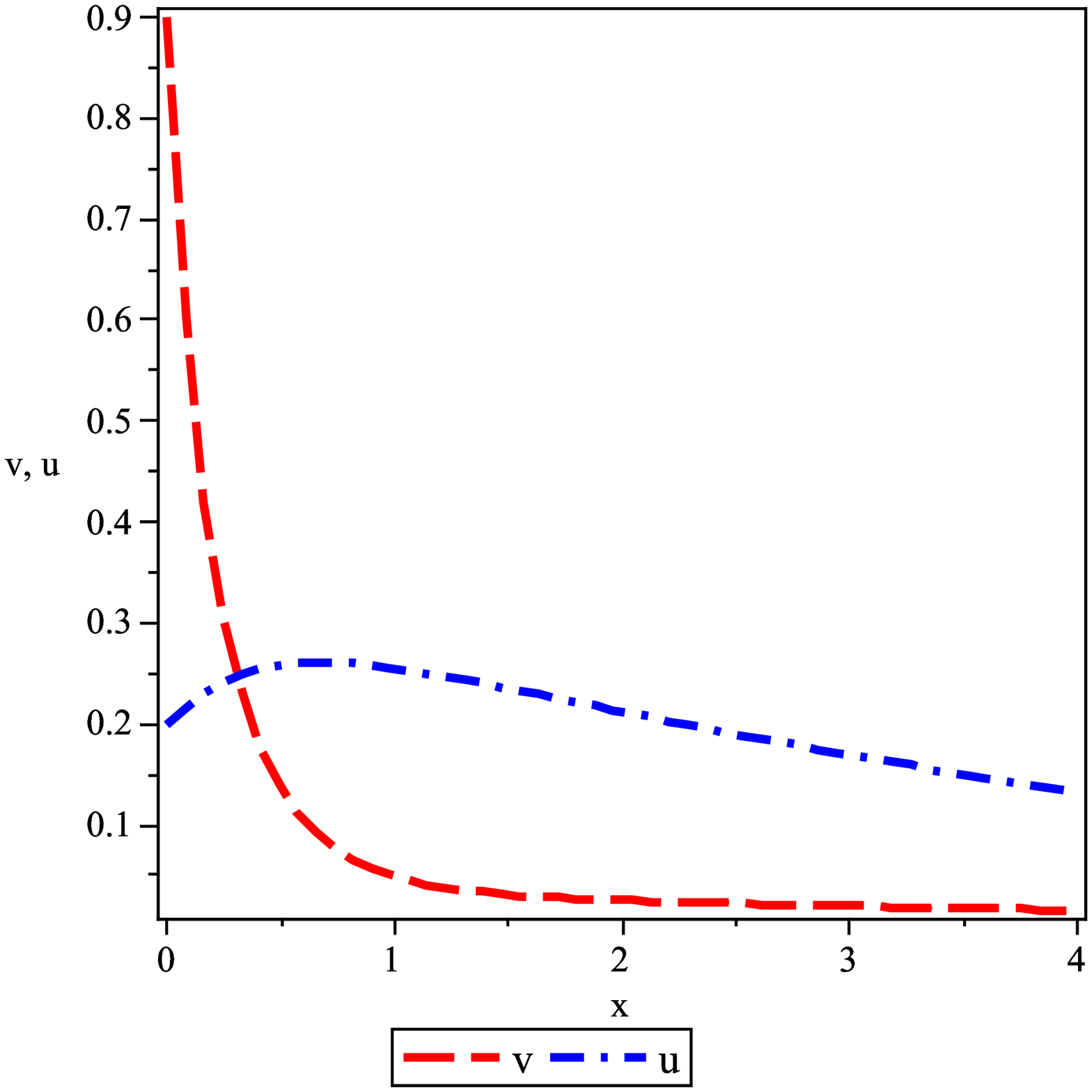}~~~~~~~~\includegraphics[height=2in]{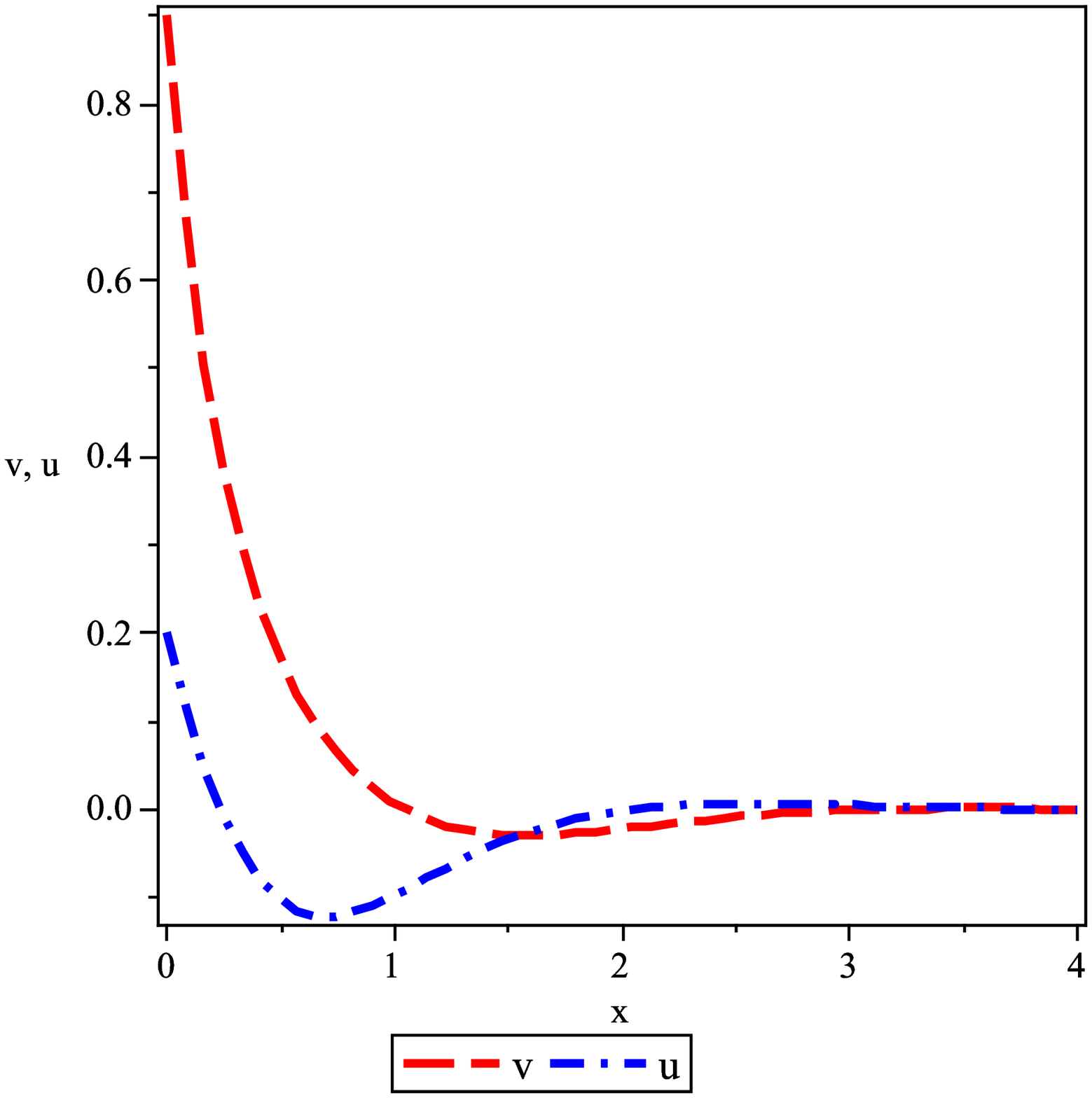}~~~~~\\
\vspace{1mm}
~~~~~~~~~~~~Fig. 14~~~~~~~~~~~~~~~~~~~~~~~~~~~~~~~~~~~~~~~Fig. 15~~~~~~~~~~~~~~~~~~~~~~~~~~~~~~~~~~~~~Fig. 16~~~~~~~~~~~\\

\vspace{1mm}

\includegraphics[height=2in]{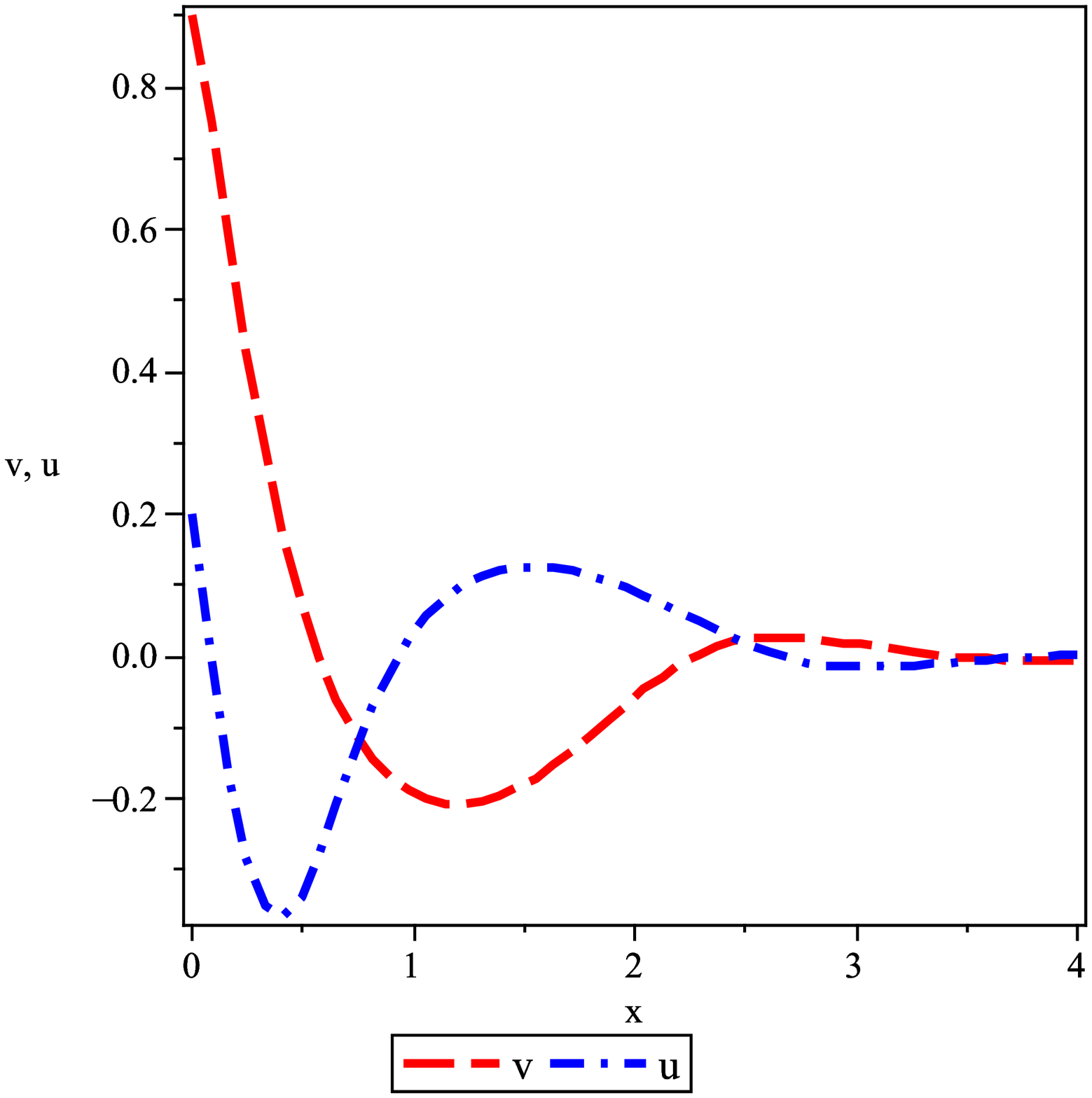}~~~~~~~~\includegraphics[height=2in]{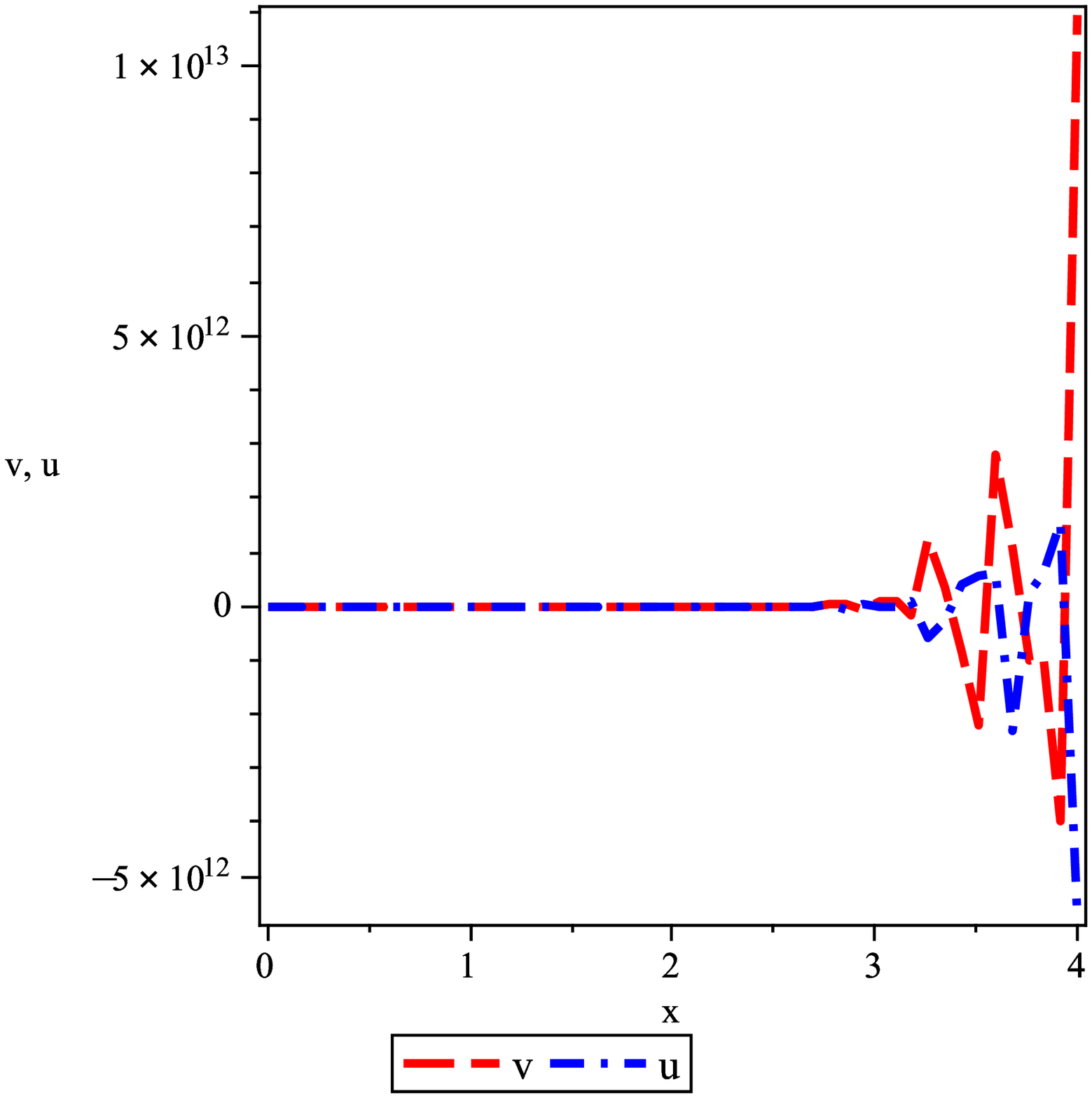}~~~~~~~~\includegraphics[height=2in]{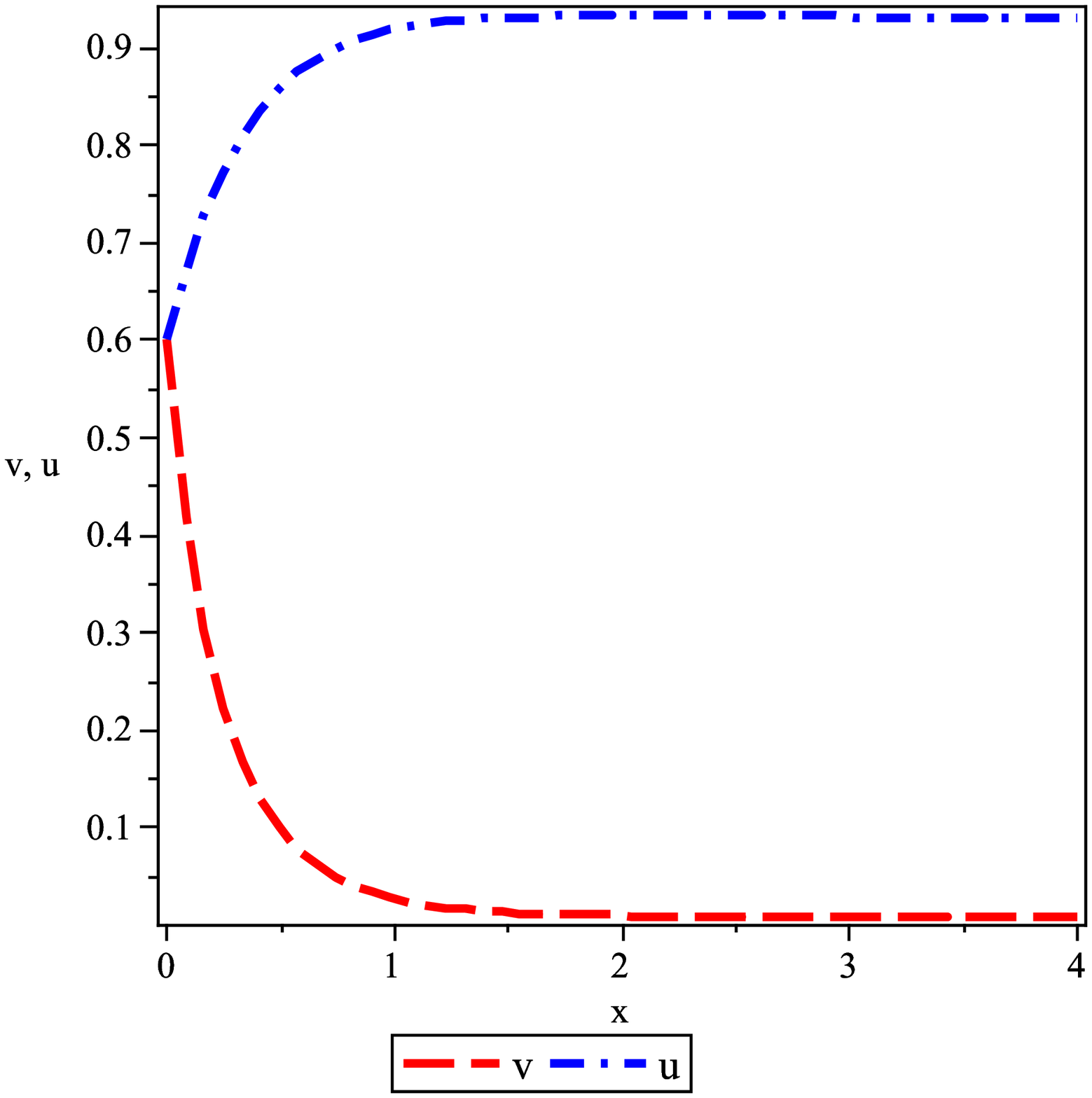}~~~~~~\\
\vspace{1mm}
~~~~~~~~~~~~Fig. 17~~~~~~~~~~~~~~~~~~~~~~~~~~~~~~~~~~~~~~~Fig. 18~~~~~~~~~~~~~~~~~~~~~~~~~~~~~~~~~~~~~Fig. 19~~~~~~~~~\\

\vspace{1mm}

\includegraphics[height=2in]{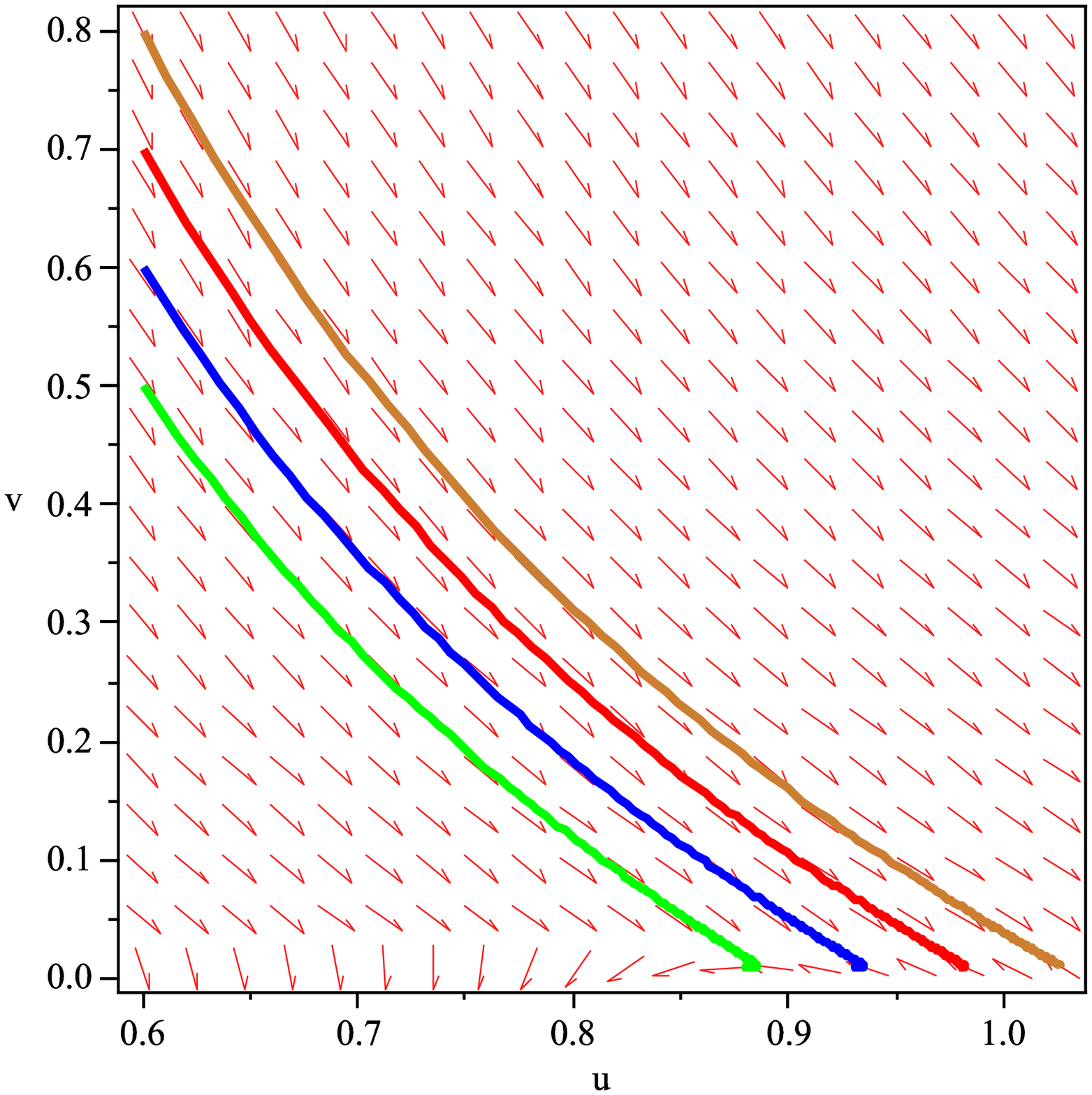}~~~~~~~~\includegraphics[height=2in]{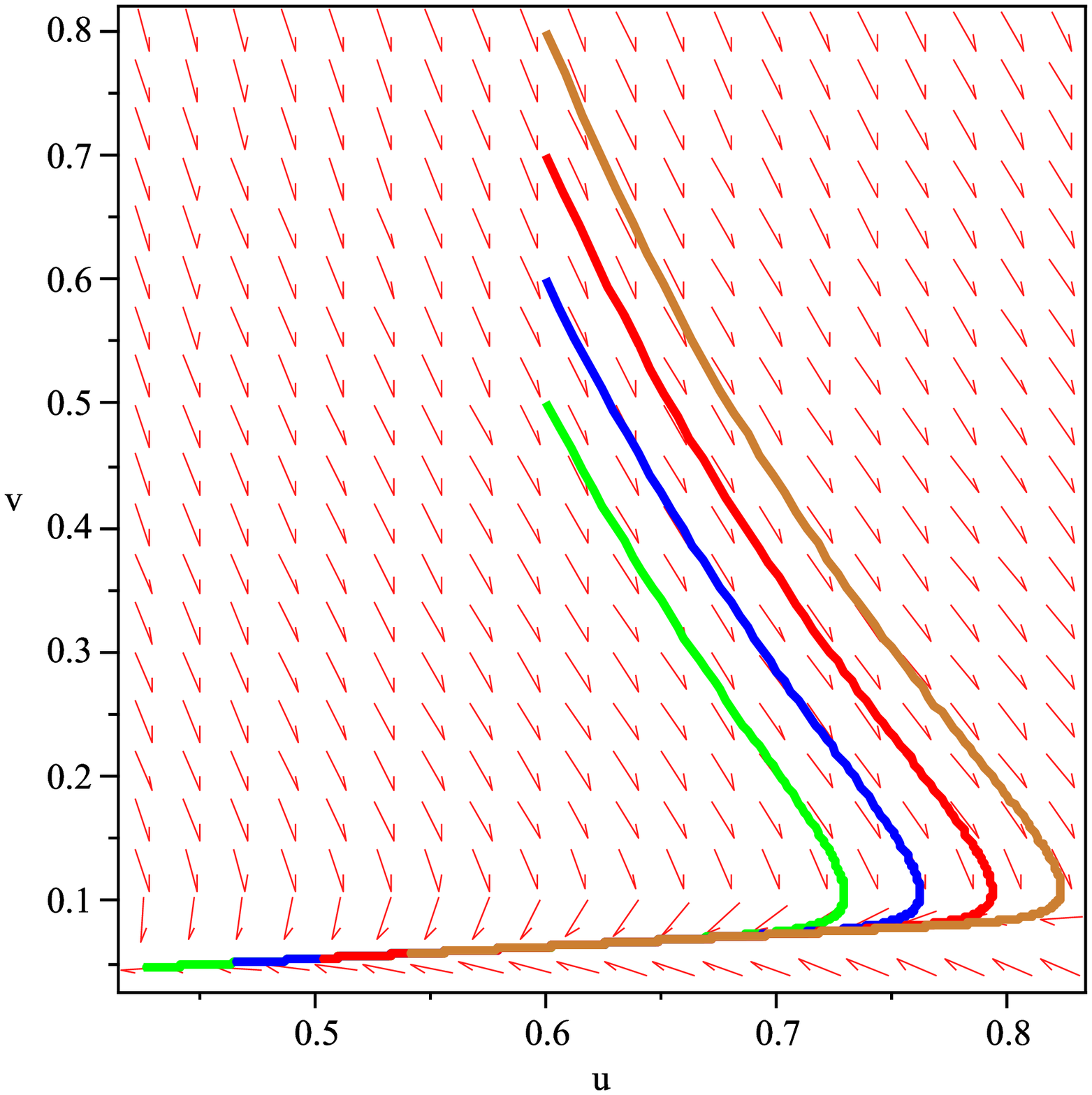}~~~~~~~~\includegraphics[height=2in]{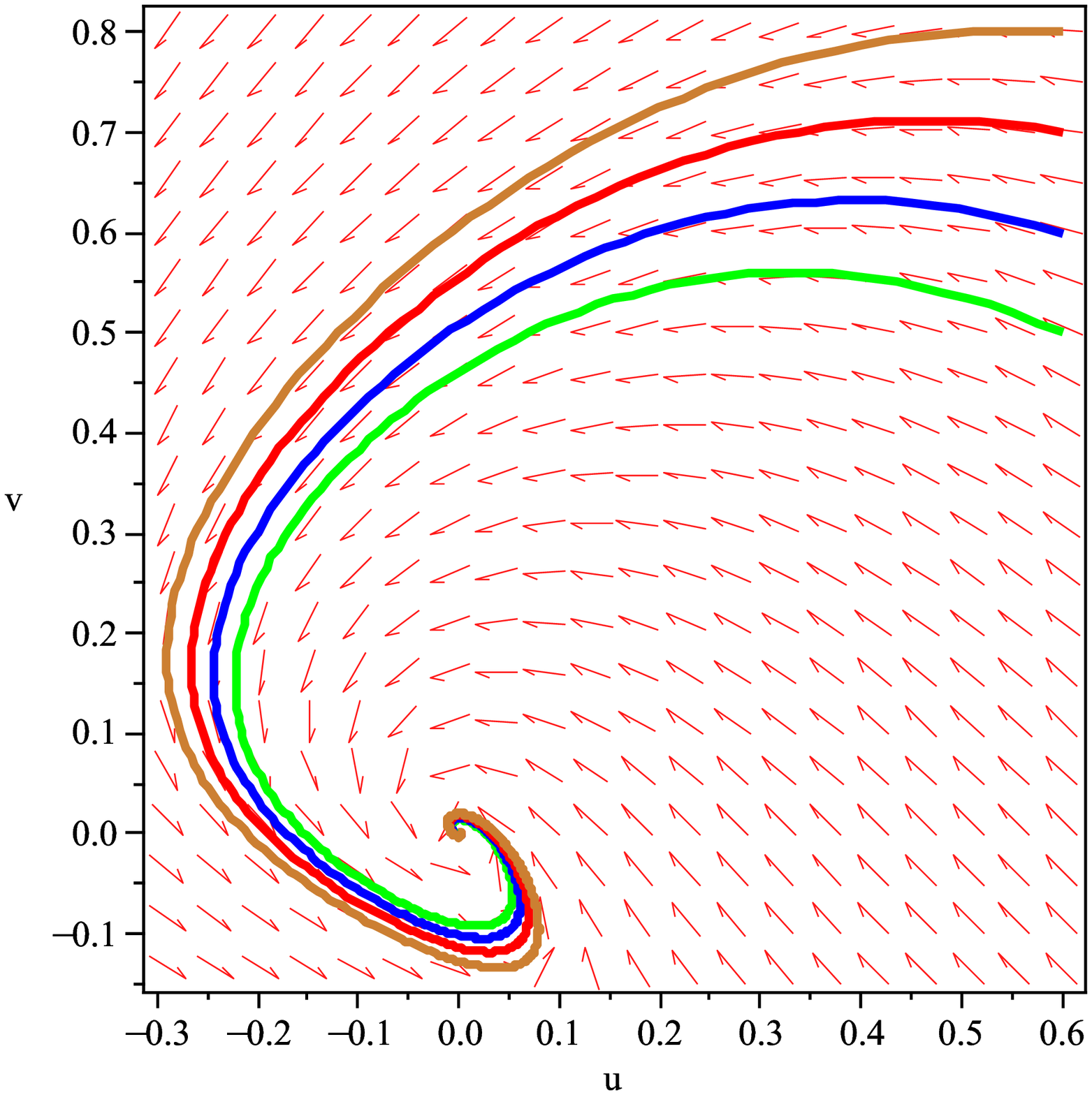}~~~~\\
\vspace{1mm}
~~~~~~~~~~~Fig. 20~~~~~~~~~~~~~~~~~~~~~~~~~~~~~~~~~~~~~~~Fig. 21~~~~~~~~~~~~~~~~~~~~~~~~~~~~~~~~~~~~~Fig. 22~~~~~~~\\

\vspace{1mm}

\includegraphics[height=1.5in]{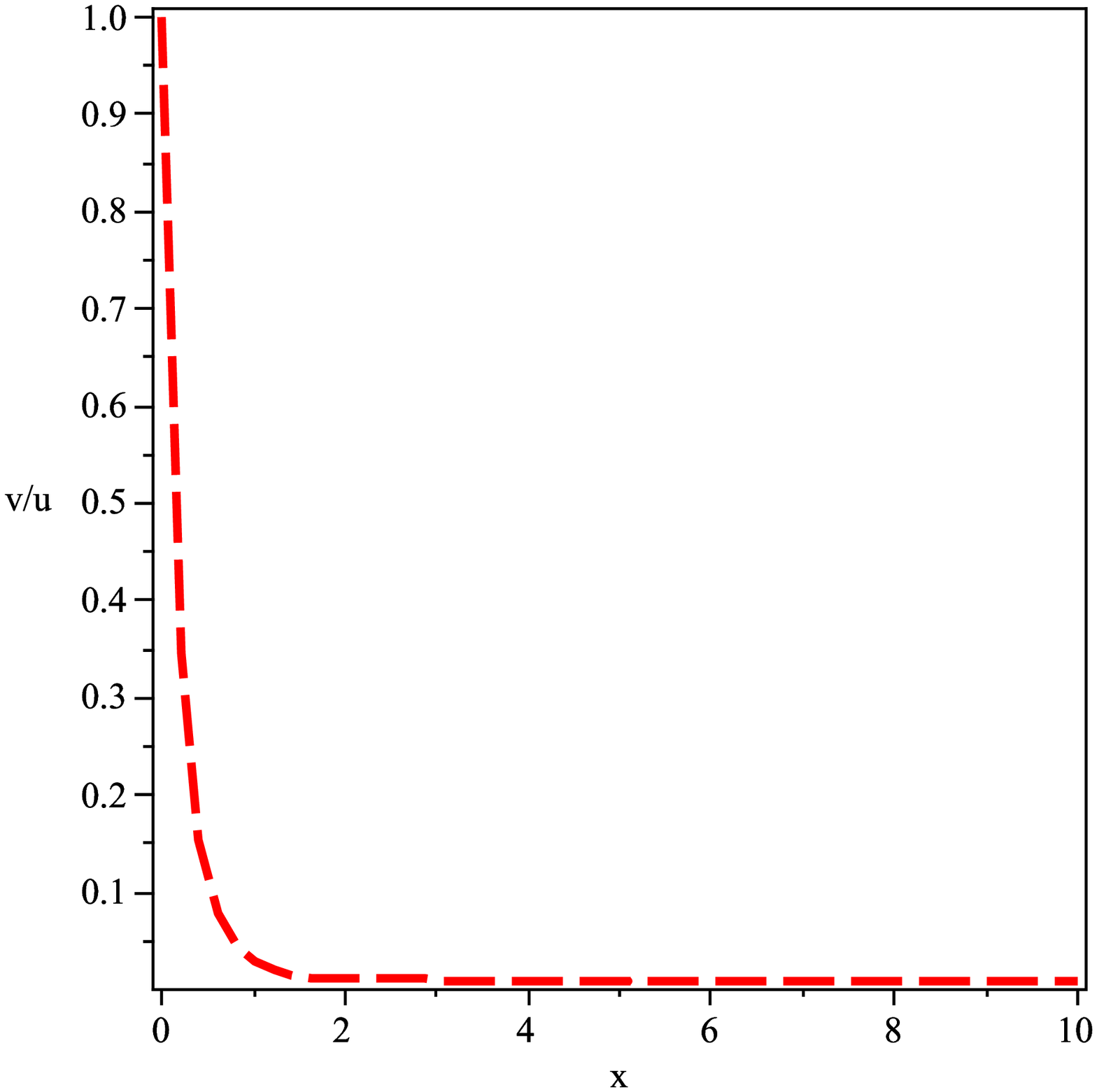}~~~~~~~~\includegraphics[height=1.5in]{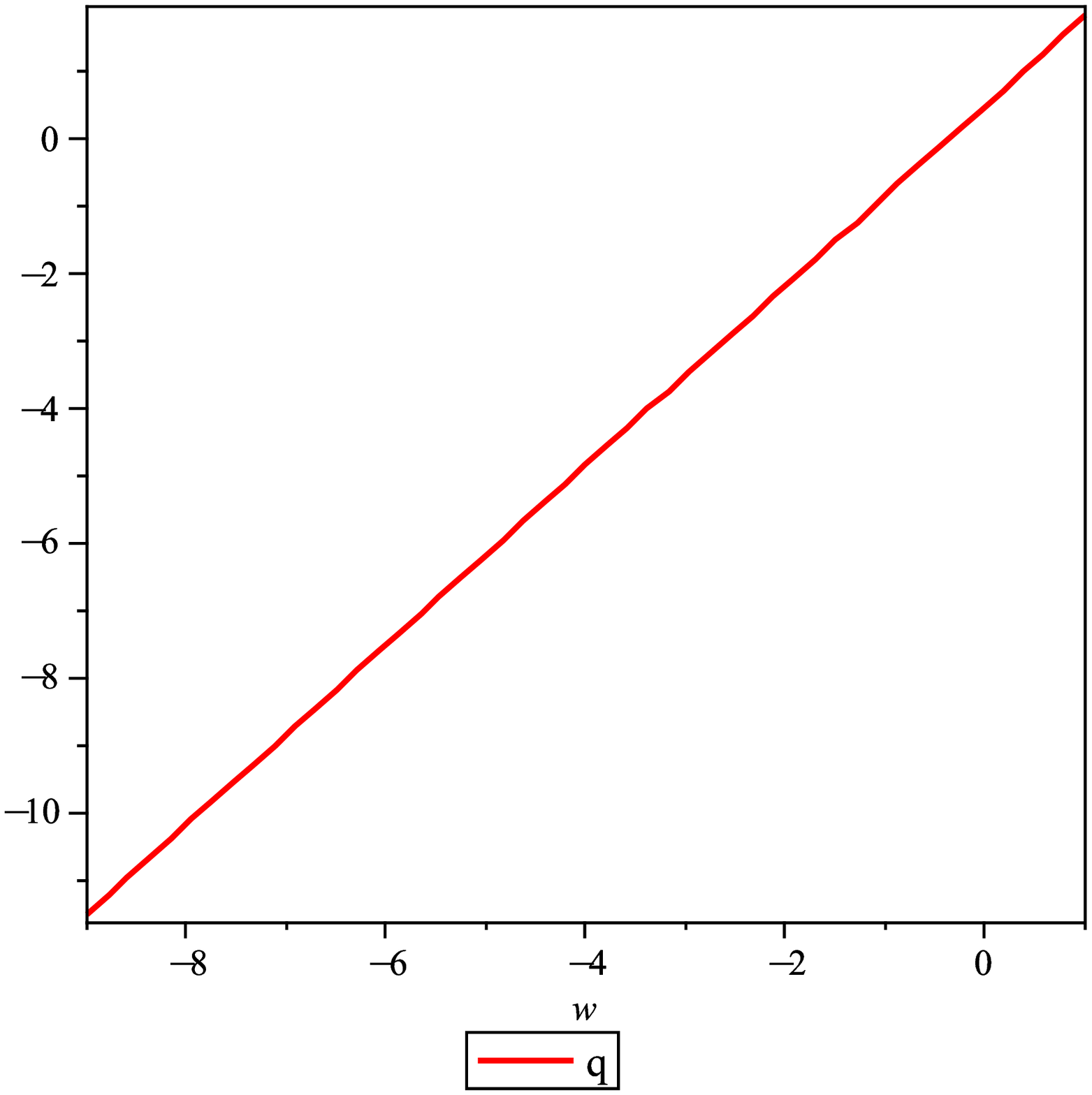}~~~~~~~~\includegraphics[height=1.5in]{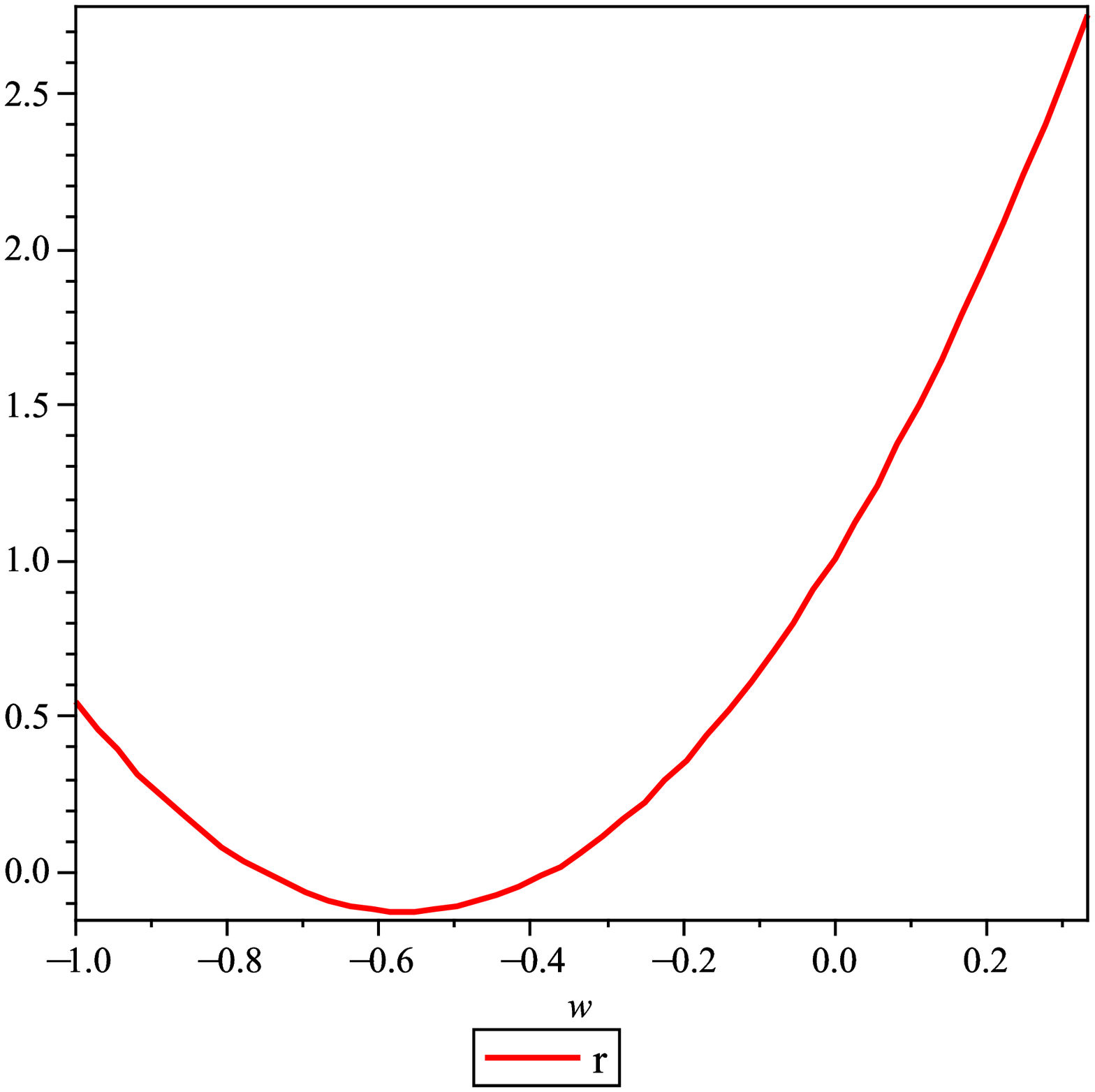}~~~~~~~\includegraphics[height=1.5in]{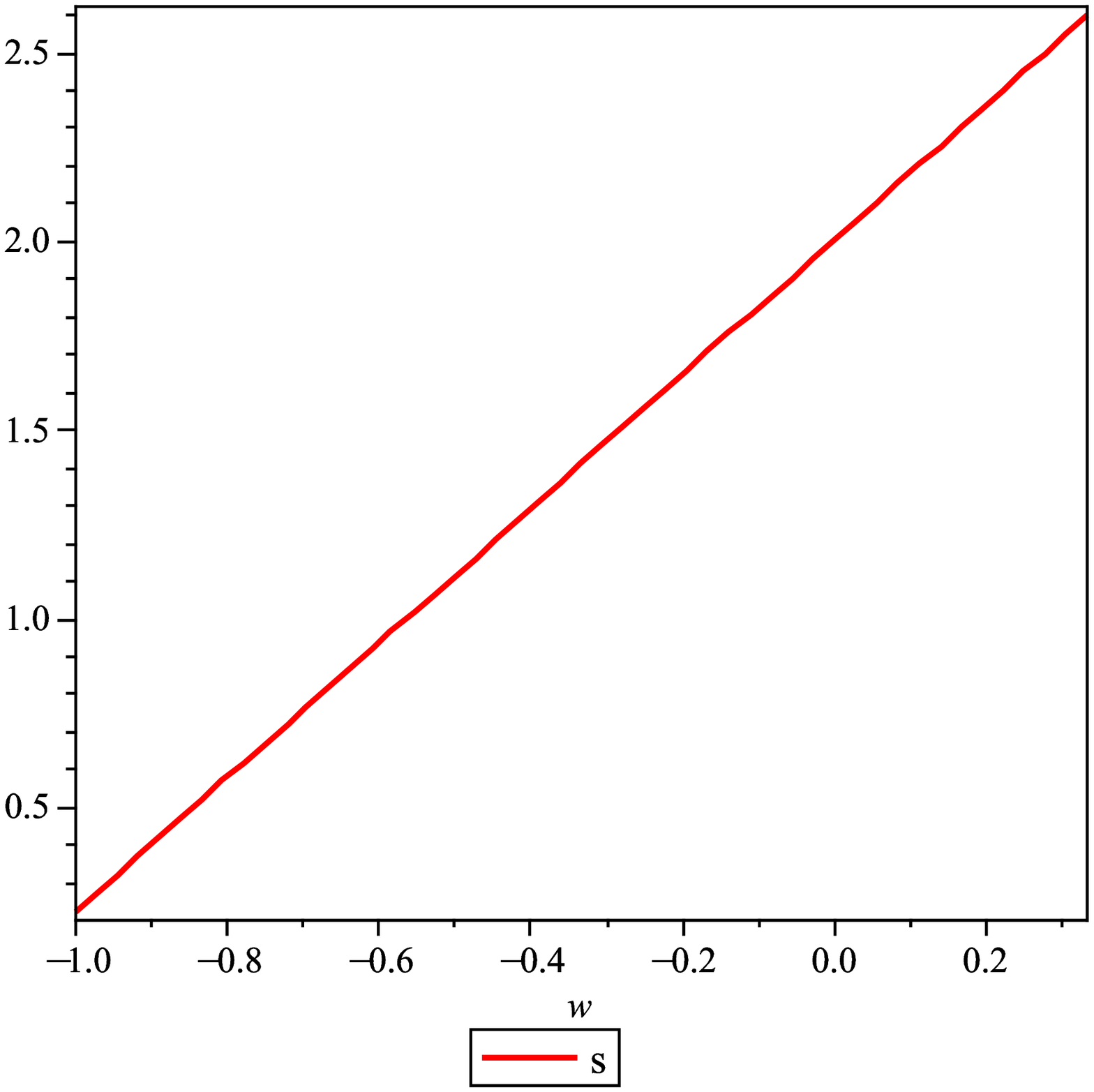}~\\
\vspace{1mm}
~~~~~~Fig. 23~~~~~~~~~~~~~~~~~~~~~~~~Fig. 24~~~~~~~~~~~~~~~~~~~~~~~~~~~~~Fig. 25~~~~~~~~~~~~~~~~~~~~~~~~~~Fig. 26\\

\vspace{1mm}

\end{figure}

\newpage

\vspace{1mm}

{\bf Description of figures:} \vspace{2mm}

Figs.14, 15, 16, 17, 18 : The dimensionless density parameters are
plotted against e-folding time for different values of
interactions, $b$. The initial condition is $v(0)=0.9, u(0)=0.2$.
The other parameters are fixed at $w=-1, C=-1$ and $r_{c}=10$.
The interactions are respectively $b=0.01, 0.1, 0.5, 1$ and $10$.\\

Fig.19: The dimensionless density parameters are plotted against
e-folding time for same initial condition. The initial condition
is $v(0)=0.6, u(0)=0.6$. The other parameters are fixed at $w=-1,
C=-1$, $r_{c}=10$ and $b=0.01$.\\

Figs.20, 21, 22 : The phase diagram of the parameters depicting an
attractor solution are obtained for different values of
interactions. The initial conditions chosen are $v(0)=0.5,
u(0)=0.6$ (green); $v(0)=0.6, u(0)=0.6$ (blue); $v(0)=0.7,
u(0)=0.6$ (red); $v(0)=0.8, u(0)=0.6$ (brown). Other parameters
are fixed at The other parameters are fixed at $w=-1, C=-1$,
$r_{c}=10$ and $b=0.01$. The interactions are respectively $b=0.01,~ 0.1$ and $1$. \\

Fig. 23 : The ratio of density parameters is shown against
e-folding time. The initial conditions chosen are v(0)=0.6,
u(0)=0.6. The other parameters are fixed at The other parameters
are fixed at $w=-1, C=-1$,
$r_{c}=10$ and $b=0.01$.\\

Fig. 24 :The deceleration parameter is plotted against the EoS
parameter. Other parameters are fixed at $\psi_{(RSII)}=-0.5$.\\

Figs.25, 26 : The statefinder parameter $r$ and $s$ are plotted
against the EoS parameter. Other parameters are fixed at $b=0.1, r_{c}=10$ and $\sigma=0.001$.\\

\subsubsection{\bf NATURE OF COSMOLOGICAL PARAMETERS}

\vspace{1mm}

{\bf 1. Deceleration Parameter: }

\vspace{2mm}

\noindent

We calculate the deceleration parameter ~$q=-1-(\dot H/H^2)$, in
this model as,
\begin{equation}
q^{(DGP)}=-1+\frac{3}{2}\left[\frac{4r_c^2\left(1+\omega_{gccg}^{(DGP)}\frac{\rho_{gccg}}{\rho}\right)}{\sqrt{4r_c^2+3\sigma}\left(\sqrt{4r_c^2+3\sigma}+\epsilon\sqrt{3\sigma}\right)}\right]
\end{equation}
where $\sigma=\frac{1}{\rho}$~. The above deceleration parameter
can be written in terms of dimensionless density parameter
$\Omega_{mcg}=\frac{\rho_{mcg}}{\rho}$ as
\begin{equation}
q^{(DGP)}=-1+\frac{3}{2}\left[\frac{4r_c^2\left(1+\omega_{gccg}^{(DGP)}\Omega_{gccg}\right)}{\sqrt{4r_c^2+3\sigma}\left(\sqrt{4r_c^2+3\sigma}+\epsilon\sqrt{3\sigma}\right)}\right]
\end{equation}
Now like the previous case we obtain
$\Omega_{mcg}=\frac{\rho_{mcg}}{\rho}=\frac{u}{u+v}$. So from the
previous equation we get,
\begin{equation}
q^{(DGP)}=-1+\frac{3}{2}\left[\frac{4r_c^2\left(1+\omega_{gccg}^{(DGP)}\frac{u}{u+v}\right)}{\sqrt{4r_c^2+3\sigma}\left(\sqrt{4r_c^2+3\sigma}+\epsilon\sqrt{3\sigma}\right)}\right]
\end{equation}
We consider the first stable critical point. At the critical point
$\left(u,v\right)\rightarrow(u_{c},v_{c})$. Hence using equation
(50) we get
\begin{equation}
q_c^{(DGP)}=-1+\frac{3}{2}Z_{(DGP)}~~,~~where~~
Z_{(DGP)}=\frac{4r_c^2\left(1+\omega_{gccg}^{(DGP)}\frac{u_{c}}{u_{c}+v_{c}}\right)}{\left\{\sqrt{4r_c^2+3\sigma}\left(\sqrt{4r_c^2+3\sigma}+\epsilon\sqrt{3\sigma}\right)\right\}}.
\end{equation}

Moreover we see that for $\sigma=0$,~~i.e., for
$\rho\rightarrow\infty$, we retrieve the results for Einstein
gravity, as given below,
\begin{equation}
q_{EG}=-1+\frac{3}{2}\left(1+\omega_{gccg}^{(DGP)}\Omega_{gccg}\right)
\end{equation}

Considering the DGP(-) model we get,

\begin{equation}
q_c^{(DGP)}=-1+\frac{3}{2}Z_{(DGP)}~~,~~where~~
Z_{(DGP)}=\frac{4r_c^2\left(1+\omega_{gccg}^{(DGP)}\frac{u_{c}}{u_{c}+v_{c}}\right)}{\left\{\sqrt{4r_c^2+3\sigma}\left(\sqrt{4r_c^2+3\sigma}-\sqrt{3\sigma}\right)\right\}}.
\end{equation}

\noindent

If we do not want to spoil the successes of the ordinary
cosmology, we have to assume the $r_{c}$ is of the order of the
present Hubble scale $H_{0}^{-1}$ \cite{Def2}. Hence $r_{c}\neq0$.
In the previous model we have seen that for $Z_{(DGP)}=0$, ~$q=-1$
and the recent cosmic acceleration is effectively realized. From
equation (53), we see that
$\left(1+\omega_{gccg}^{(DGP)}\frac{u_{c}}{u_{c}+v_{c}}\right)=0$,
since $r_{c}\neq0$. From this relation we get,
\begin{equation}
\omega_{gccg}^{(DGP)}=-\frac{u_{c}+v_{c}}{u_{c}}
\end{equation}
Now since $\frac{u_{c}+v_{c}}{u_{c}}>1$, we should have,
\begin{equation}
\omega_{gccg}^{(DGP)}<-1
\end{equation}
{\bf The above range for the EoS parameter indicates the phantom
era for GCCG type DE}.\\
Hence it is evident that when
$\omega_{gccg}^{(DGP)}=-\frac{u_{c}+v_{c}}{u_{c}}$,~~~$Z_{(DGP)}=0$,
and hence ~$q=-1$. This is consistent with the recent cosmic
acceleration.

\noindent

Moreover the Hubble parameter can be obtained as,
\begin{equation}
H=\frac{2}{3Z_{(DGP)}t},
\end{equation}
where we have ignored the integration constant. Integration of
eqn.(56) yields
\begin{equation}
a(t)=a_0t^{\frac{2}{3Z_{(DGP)}}},
\end{equation}
which gives a power law form of the expansion. In order to realize
the accelerating scenario of the universe, we should have
$\frac{2}{3Z_{(DGP)}}>1~ i.e.,~0<Z_{(DGP)}<\frac{2}{3}$. Using
this range of $Z_{(DGP)}$ in the equation
$q_{c}^{(DGP)}=-1+\frac{3}{2}Z_{(DGP)}$. We get the range of
$q_{c}^{(DGP)}$ as $-1<q_{c}^{(DGP)}<0$. Therefore the
deceleration parameter is negative and hence the result is
consistent with the fact that the universe is undergoing an
accelerated expansion.\\

{\bf 2. Statefinder Parameters:}

In the DGP brane model, we have the following expressions for the
statefinder parameters, $r$ and $s$ as given below,
\begin{equation}
r_{(DGP)}=\left(1-\frac{3Z_{(DGP)}}{2}\right)\left(1-3Z_{(DGP)}\right).
\end{equation}
and
\begin{equation}
s_{(DGP)}=Z_{(DGP)}.
\end{equation}

\section{DETAILED GRAPHICAL STUDY OF PHASE PLANE ANALYSIS}

\noindent

Figs. 1 to 5 and 14 to 18, shows the plots of density parameters
$u$ and $v$, respectively for RS II and DGP brane model. We see
that in case of RS II brane as the brane tension $\lambda$
decreases (Figs. 1 to 4), more and more irregularity creep in, as
far as the DE density parameter $u$ is concerned. But all the four
figures show an energy dominated universe, consistent with
observational data. In fig. 5, with a larger value of interaction,
we get a matter dominated universe (unphysical situation), which
really indicates that the interaction coupling parameter, $b$
should be a small positive value. In case of DGP brane, we see
that with the increase in interaction between energy and matter,
the density parameters become more and more comparable to each
other, giving a solution to the cosmic coincidence problem . An
identical result was obtained for GCCG in LQC in
\cite{Chowdhury1}. Figs. 6 and 19, shows almost the same results
for identical scenario. In figs. 7, 8 and 9, Phase diagrams for RS
II brane have been obtained. The figs. 7 and 8 have been generated
for different values of brane tension $\lambda$. It is seen that
as the tension decreases there is a greater tendency of the
solution moving towards an attractor, thus giving a perfect
attractor solution. In fig. 9 with an higher interaction, we get a
far better attractor solution, but the direction of flow is
reversed. This can be attributed to the fact that as the
interaction grows in magnitude DE interferes more and more with
DM, until the matter loses its dominance and its place is taken by
the energy, thus giving a perfectly energy dominated scenario.
This shift of power may be responsible for many unexplained
phenomena of cosmology including the present one. In the figs. 20,
21 and 22, we have obtained the phase diagrams for the DGP brane
model, with gradually increased values of interaction. Just like
the previous model we see that with the increase in interaction,
there is a greater tendency of the flow going towards a specific
attractor point. In the figs. 10 and 23, plots for the ratio of
the density parameters have been generated against the e-folding
time, for the two models respectively. In the plots, it is evident
that $\frac{v}{u}$ decreases with time, thus exhibiting an energy
dominated scenario.

Figs. 11 and 24 shows the plot of deceleration parameter, $q$
against the EoS parameter, for the two models respectively. In the
fig. 11, $q$ remain in the negative level throughout the
quintessence era thus exhibiting the recent cosmic acceleration
for the RS II model. The only condition being the negativity of
the brane tension, $\lambda$. But in the fig. 24, we see that $q$
remains in the negative level for $\omega<-1$, i.e. only in the
phantom region. This not only shows that GCCG is a DE fluid with
far lesser negative pressure compared to other dark energy models,
but also it shows that the combination of GCCG in DGP brane model
gives an inferior model of the universe compared to the other
combinations, like MCG in DGP brane (refer to \cite{Rudra1}). The
statefinder parameters have been plotted in the figures 12 and 13
for RS II brane and in figs. 25 and 26 for the DGP brane. In the
figs. 12 and 13, we see that $r$ decreases whereas $s$ increases
with the increase in EoS parameter. But in case of DGP brane, $r$
initially decrease, and then increase after reaching a minimum
value. $s$ increases with the increases in EoS parameter just like
the previous model.

\section{STUDY OF FUTURE SINGULARITY}

\noindent

We speculate that any energy dominated model of the universe
undergoing an accelerated expansion will result in a future
singularity. The study of dynamics of an accelerating universe in
the presence of DE and DM is in fact incomplete without the study
of these singularities, which are the ultimate fate of the
universe.  It is known that the universe dominated by phantom
energy ends with a future singularity known as Big Rip
\cite{Caldwell1}, due to the violation of dominant energy
condition (DEC). But other than this there are other types of
singularities as well. Nojiri et al \cite{Nojiri1} studied the
various types of singularities that can result from a phantom
energy dominated universe. These possible singularities are
characterized by the growth of energy and curvature at the time of
occurrence of the singularity. It is found that near the
singularity quantum effects becomes very dominant which may
alleviate or even prevent these singularities. So it is extremely
necessary to study these singularities and classify them
accordingly so that we can search for methods to eliminate them.
The appearance of all four types of future singularities in
coupled fluid dark energy, $F(R)$ theory, modified Gauss-Bonnet
gravity and modified $F(R)$ Horava-Lifshitz gravity was
demonstrated in \cite{Nojiri2}. The universal procedure for
resolving such singularities that may lead to bad phenomenological
consequences was proposed. In Rudra et al \cite{Rudra1} it has
been shown that in case of MCG in Brane-world, both Type I and
Type II singularities are possible. In Chowdhury et al
\cite{Chowdhury1} it was shown that in case of GCCG in LQC, the
universe is absolutely free from any type of singularities. We
proceed to study the singularities for the present case:

\subsection{TYPE I Singularity (Big Rip singularity)}
If $\rho\rightarrow\infty$ , $|p|\rightarrow\infty$ when
$a\rightarrow\infty$ and $t\rightarrow t_{s}$. Then the
singularity formed is said to be the Type I singularity.

In the present case by considering the GCCG equation of state from
equation (1) we find that there is no possibility for TypeI
singularity, i.e., Big Rip singularity, since $\alpha>0$. This is
in absolute accordance with P. F. Gonz´alez-Diaz who has
successfully shown that by considering GCCG as the DE, Big Rip can
easily be avoided, thus giving a singularity free late universe.

\subsection{TYPE II Singularity (Sudden singularity)}
If $\rho\rightarrow\rho_{s}$ and $\rho_{s}\sim0$, then
$|p|\rightarrow-\infty$ for $t\rightarrow t_{s}$ and $a\rightarrow
a_{s}$, then the resulting singularity is called the Type II
singularity.

In this case we consider the equation of state for GCCG, like the
previous case for our investigation. We see that if
$\rho\rightarrow\rho_{s}$ and $\rho_{s}\sim0$, then
$|p|\rightarrow0$ for $t\rightarrow t_{s}$ and $a\rightarrow
a_{s}$. Hence there is no possibility of the type II singularity
or the sudden singularity in case of GCCG, primarily because
$\alpha>0$ and $w<0$.

\subsection{TYPE III Singularity}
For $t\rightarrow t_{s}$, $a\rightarrow a_{s}$,
$\rho\rightarrow\infty$ and $|p|\rightarrow\infty$. Then the
resulting singularity is Type III singularity. It is quite evident
from the equation of state of GCCG that it does not support this
type of singularity.

\subsection{TYPE IV Singularity}
For $t\rightarrow t_{s}$, $a\rightarrow a_{s}$, $\rho\rightarrow0$
and $|p|\rightarrow0$. Then the resulting singularity is Type IV
singularity. This type of singularity is not supported by GCCG
type DE.

As a remark, one should stress that our consideration is totally
classical. Nevertheless, it is expected that quantum gravity
effects may play significant role near the singularity. It is
clear that such effects may contribute to the singularity
occurrence or removal too. Unfortunately, due to the absence of a
complete quantum gravity theory only preliminary estimations may
be done.

\section{CONSEQUENCES OF THE EXPANDING UNIVERSE: DISTANCE MEASUREMENT OF THE UNIVERSE}

\noindent

Cosmography is a field where we are concerned with the measurement
of the Universe. In fact there are many ways to specify the
distance between two points. This is primarily because, in the
expanding and accelerating Universe, the distances between
co-moving objects are constantly changing, and Earth-bound
observers look back in time as they look out in distance. The
unifying aspect is that all distance measures somehow measure the
separation between events on radial null trajectories, i.e.,
trajectories of photons which terminate at the observer. Here we
will compute and discuss various cosmological distance measures
such as the look-back time, luminosity distance, proper distance,
angular diameter distance, co-moving volume, distance modulus and
probability of intersecting objects.

\subsection{LOOK-BACK TIME}

As light travels with finite speed, it takes time for it to cover
the distance related to the redshift it encountered. The
difference between the age of the Universe now (at observation)
and the age of the Universe at the time the photons were emitted
(according to the object) is defined as the Lookback time to an
object. So, a look into space is always a look back in time. It is
used to predict properties of high redshift objects with
evolutionary models, such as passive stellar evolution for
galaxies. Thus if a photon emitted by a source at the instant $t$
and received at the time $t_{0}$ then the photon travel time or
the lookback time $t_{0} - t$ is defined by \cite{lb1}

\begin{equation}
t_{0}-t=\int_{a_{0}}^{a}\frac{da}{\dot{a}}
\end{equation}

where $a_{0}$ is the present value of the scale factor of the
universe and can be obtained from (57) at $t = t_{0}$. The
redshift is an important observable parameter as they can be
measured easily from the spectral lines, and the redshift
increases with the recession of the object from us. Look-back time
is used to predict properties of high-redshift objects with
evolutionary models, such as passive stellar evolution for
galaxies. The redshift z can be defined by

\begin{equation}
\frac{a_{0}}{a}=1+z=\left(\frac{t_{0}}{t}\right)^{\frac{2}{3Z}}
\end{equation}
which gives the look-back time in the following form
\begin{equation}
t-t_{0}=\frac{2}{3ZH_{0}}\left\{\frac{1}{\left(1+z\right)^\frac{3Z}{2}}-1\right\}
\end{equation}

For accelerating universe we have already get $Z <\frac{2}{3}$ .
Early universe is represented by $z\rightarrow\infty$ implies
$t\rightarrow0$ and late universe $z\rightarrow-1$, which
equivalently implied $t\rightarrow\infty$. Also $z\rightarrow0$
gives the present age $t\rightarrow t_{0}$ of the universe.

\subsection{PROPER DISTANCE}

\noindent

We know that light needs time to get from an object to the
observer. Therefore we can define a distance that may be measured
between the observer and the object with a ruler at the time the
light was emitted, as the proper distance. When a photon emitted
by a source at time $t_{0}$ and received by an observer at time
$t$ then the proper distance between them is defined by \cite{lb1,
lb2}.

\begin{equation}
d=a_{0}\int_{a}^{a_{0}}\frac{da}{a\dot{a}}=a_{0}\int_{t}^{t_{0}}\frac{dt}{a}
\end{equation}
which gives
\begin{equation}
d=\frac{2}{H_{0}\left(3Z-2\right)}\left\{1-\frac{1}{\left(1+z\right)^{\frac{3Z}{2}-1}}\right\}
\end{equation}

As far as the current epoch is concerned the proper distance may
also called the co-moving distance (line of sight) of the
Universe. So between two nearby objects in the Universe, the
distance between them remains constant with epoch if the two
objects are moving with the Hubble flow. In other words, it is the
distance between them which would be measured with rulers at the
time they are being observed divided by the ratio of the scale
factor of the Universe ($a$ at that time to $a$ now). Next thing
to be defined is the transverse co-moving distance, which is a
quantity used to get the co-moving distance perpendicular to the
line of sight. For flat universe, the transverse co-moving
distance is always identical to the co-moving distance (line of
sight). That means the transverse co-moving distance = proper
distance = d, for any model following the power law form of
expansion.

\subsection{LUMINOSITY DISTANCE}

\noindent

If $L$ be the total energy emitted by the source per unit time and
$\ell$ be the apparent luminosity of the object then the
luminosity distance is defined by \cite{lb1, lb2}

\begin{equation}
d_{L}=\left(\frac{L}{4\pi\ell}\right)^{\frac{1}{2}}=d\left(1+z\right)=\frac{2}{H_{0}\left(3Z-2\right)}\left\{\left(1+z\right)-\frac{1}{\left(1+z\right)^{\frac{3Z}{2}}}\right\}
\end{equation}

\subsection{ANGULAR DIAMETER DISTANCE}

\noindent

We define the angular diameter of a light source of proper
distance $D$ observed at $t_{0}$ by \cite{lb1, lb2}
\begin{equation}
\delta=\frac{D\left(1+z\right)^{2}}{d_{L}}
\end{equation}
Now the ratio of the source diameter to its angular diameter (in
radians) is defined as the angular diameter distance $d_{A}$ as
furnished below,

\begin{equation}
d_{A}=\frac{D}{\delta}=d_{L}\left(1+z\right)^{-2}=d\left(1+z\right)^{-1}
\end{equation}

For our models following power law form of expansion angular
diameter distance $(d_{A})$ is given by,
\begin{equation}
d_{A}=\frac{2}{H_{0}\left(3Z-2\right)}\left\{\frac{1}{1+z}-\frac{1}{\left(1+z\right)^{\frac{3Z}{2}-2}}\right\}
\end{equation}

It is used to convert angular separations in telescope images into
proper separations at the source. It is famous for not increasing
indefinitely as $z\rightarrow\infty$; it gets inverted at
$z\sim{1}$ and thereafter more distant objects actually appear
larger in angular size. The angular diameter distance is maximum
at

\begin{equation}
z_{max}=\left(\frac{2}{3Z-2}\right)^\frac{2}{3\left(2-Z\right)}-1
\end{equation}
and corresponding maximum angular diameter $d_{A}|_{max}$ taking
the form
\begin{equation}
d_{A}|_{max}=\frac{1}{H_{0}\left(3Z-2\right)}\left[2^{1+\frac{2}{3\left(2-Z\right)}}\left(\frac{1}{3Z-4}\right)^{\frac{2}{3\left(Z-2\right)}}-2\left\{4^{\frac{1}{3\left(2-Z\right)}}\left(\frac{1}{3Z-4}\right)^{\frac{2}{3\left(Z-2\right)}}\right\}^{2-\frac{3Z}{2}}\right]
\end{equation}
All the above four parameters have been plotted against the
redshift parameter, $z$ in fig. 27.

\begin{figure}
\includegraphics[height=3in]{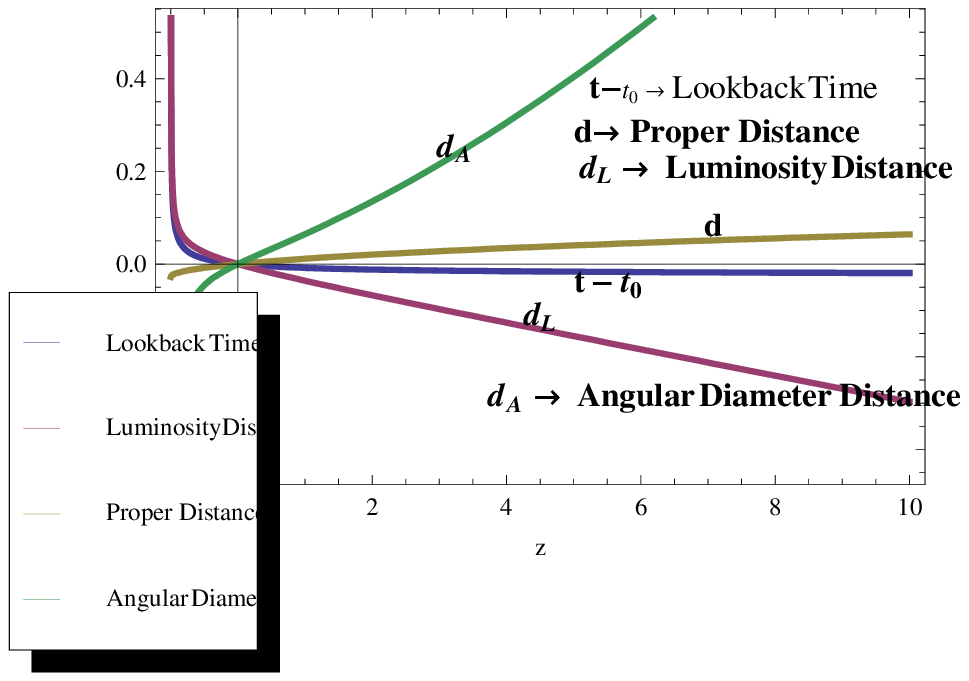}~~~~\\
\vspace{1mm}
Fig. 27~~~~\\

\end{figure}

\subsection{CO-MOVING VOLUME}

The co-moving volume $V_{C}$ is the volume measure in which number
densities of non-evolving objects locked into Hubble flow are
constant with redshift. We define it as \cite{lb1, lb2}
\begin{equation}
dV_{C}=D_{H}\frac{\left(1+z\right)^{2}d_{A}}{E(z)}d\Omega
dz=\frac{1}{H_{0}}\left(1+z\right)^{2-\frac{3Z}{2}}d_{A}^{2}d\Omega
dz
\end{equation}
where $d\Omega$ is the solid angle element and $d_{A}$ is the
angular diameter, $E(z)=\frac{H(z)}{H0}$ and $D_{H}
=\frac{c}{H_{0}}$ is the Hubble distance (c is the velocity of
light) and in our model we assume $c = 1, d
 = 1$, $H_{0}$ = 72km/s/Mpc.

So, the co-moving volume is proper volume times the ratio of scale
factors now to then to the third power. Co-moving volume element
$\frac{dV_{C}}{dz}$ are drawn in figure 28. We see that there is a
gradual increase with increase in redshift $z$.

\subsection{DISTANCE MODULUS}

\noindent

The distance modulus is define by

\begin{equation}
D_{M}=5\log{\left(\frac{d_{L}}{10pc}\right)}
\end{equation}
because it is the magnitude of difference between objects observed
bolometric (i.e., integrated over all frequencies) flux and what
it would be if it were at $10$ pc (this was once thought to be the
distance to Vega) and $d_L$ is the luminosity distance. Distance
modulus $D_{M}$ as a function of redshift have been shown in
figure 29. We see that for $z>0$, we do not get any plot for
$D_M$. But for $z<0$, $D_M$ increases as $z$ decreases.

\begin{figure}
\vspace{2mm}
 ~~~~\includegraphics[height=2in]{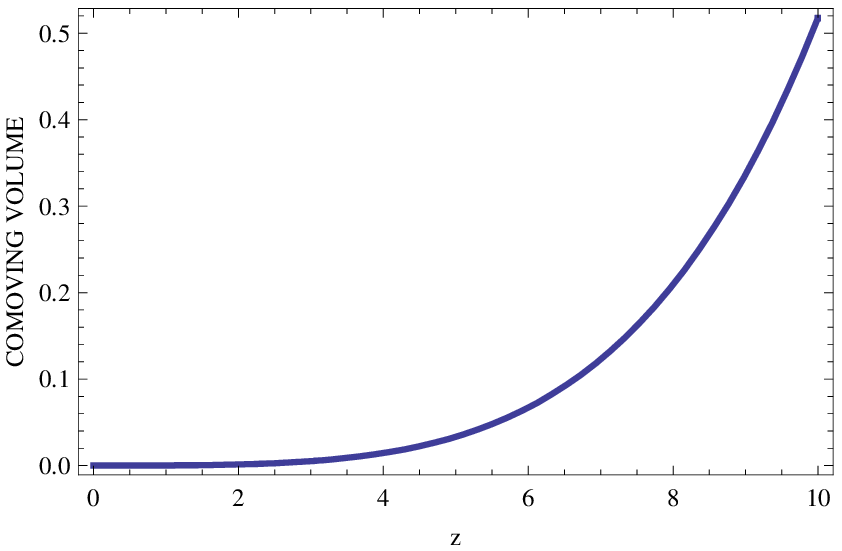}~~~~~~~\\
\vspace{2mm}
Fig. 28~~~~~\\
\vspace{1mm}

\includegraphics[height=2in]{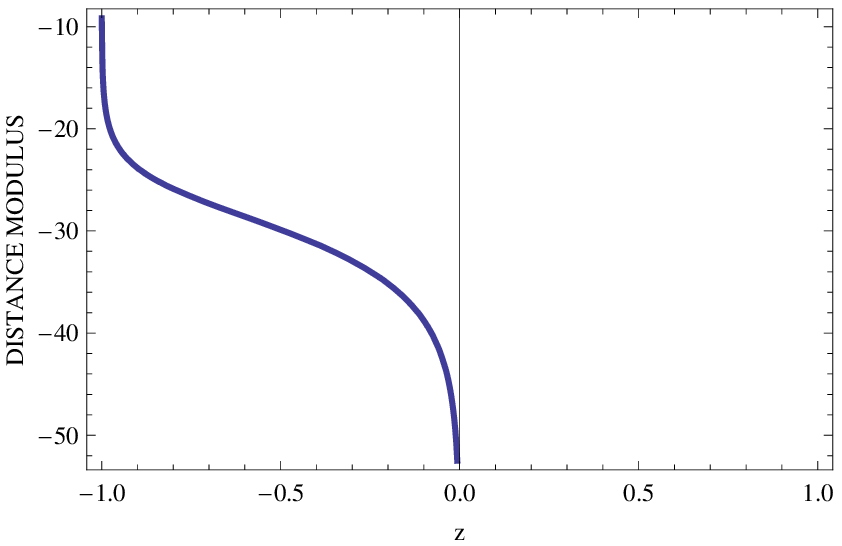}~~~~~~~\includegraphics[height=2in]{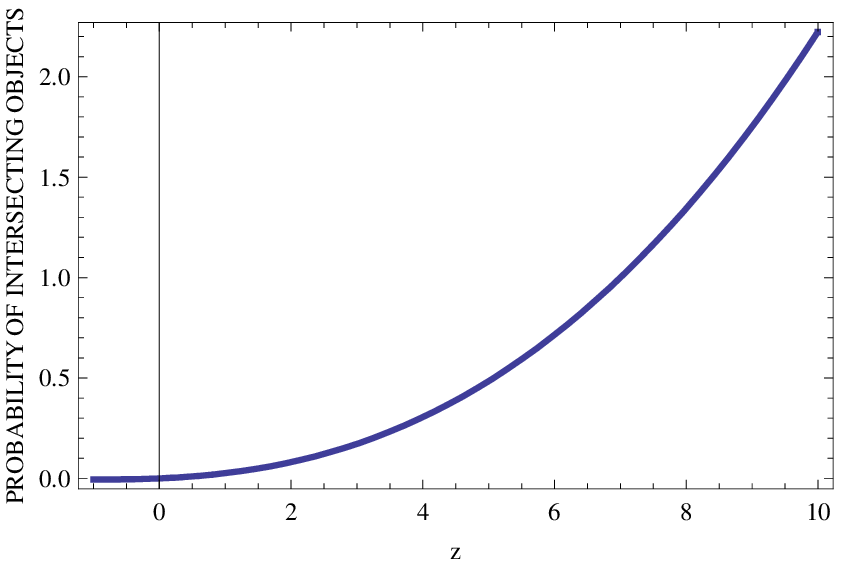}~\\
\vspace{1mm}
Fig. 29~~~~~~~~~~~~~~~~~~~~~~~~~~~~~~~~~~~~~~~~~~~~~~~~~~~~~~~~~Fig. 30~~~~~\\
\end{figure}

\subsection{PROBABILITY OF INTERSECTING OBJECTS}

It is defined as the incremental probability $dP$ that a line of
sight will intersect one of the objects in redshift interval $dz$
at redshift $z$. It is given by \cite{lb1, lb2}

\begin{equation}
dP=n(z)\sigma(z)D_{H}\frac{\left(1+z\right)^{2}}{E(z)}dz
\end{equation}
where $n(z)$ is the co-moving number density and $\sigma(z)$ areal
cross-section. Assuming $n(z)\sigma(z)=1$, we obtain
\begin{equation}
dP=\frac{1}{H_{0}}\left(1+z\right)^{2-\frac{3Z}{2}}dz
\end{equation}
For our model the expression of Probability of intersecting
objects becomes
\begin{equation}
P=\frac{2}{H_{0}\left(6-3Z\right)}\left\{\left(1+z\right)^{3-\frac{3Z}{2}}-1\right\}
\end{equation}
In figure 30 we draw intersection probability $P$ as a function of
redshift. We see that, $P$ increases as $z$ increases.

\section{CONCLUSION}

\noindent

In this work, we have considered a combination of Generalized
cosmic Chaplygin gas in standard Brane-world models. Two different
models, namely the RS II brane and the DGP brane models have been
considered for our evaluations. Our basic idea was to study the
background dynamics of GCCG in detail when it is incorporated in
brane gravity. Because of the complexity of the expressions it was
impossible to find direct solutions for the system. So we resorted
to dynamical system analysis for our computations. Dynamical
system analysis was successfully carried out, critical points were
found and the stability of the system around those critical points
was tested. Graphical analysis was done to get an explicit picture
of the outcome of the work. In order to find a solution for the
cosmic coincidence problem, a suitable interaction between DE and
DM was considered. Figures of density parameters were drawn for
different values of interaction. It was found that increase in
interaction resulted in more and more comparable values of the
density parameters of GCCG and DM. Since the tendency of DE
domination over DM is lesser in case of GCCG compared to MCG, GCCG
is identified as a dark fluid with a lesser negative pressure
compared to MCG or any other forms of DE.

It was found that GCCG in RS II brane is consistent with the late
cosmic acceleration only if the brane tension is negative. For
GCCG in DGP brane an accelerated expansion is realized only in the
phantom era of the DE. In the quintessence era there is no
possibility of an accelerating scenario. This is a very important
result as far as modern cosmology is concerned. This really shows
the less effectiveness of GCCG as a DE compared to others, like
MCG which produced an accelerating scenario in DGP brane in the
quintessence era itself. {\bf From the above results we can come
to the conclusion that although GCCG with a far lesser negative
pressure compared to other DE models, can overcome the relatively
weaker gravity of RS II brane, with the help of the negative brane
tension, yet for the DGP brane model with much higher gravitation,
the incompetency of GCCG is exposed, and it cannot produce the
accelerating scenario until and unless it reaches the phantom
era}.

The dynamical system of equations characterizing the system was
formed and a stable scaling solution was obtained. Hence this work
can be considered to be a significant one, as far as the solution
of cosmic coincidence problem is concerned. Study of future
singularities had been carried out in detail. The model was
investigated for all possible types of future singularities. From
the above analysis we conclude that the combination of GCCG in
brane gravity gives a perfect singularity free model (just like
GCCG in LQC) for an expanding universe undergoing a late
acceleration. Statefinder parameters were calculated and plots
were generated for them. The evolutionary trajectories obtained,
when compared with those of the $\Lambda$CDM model gave clear
differences, thus characterizing the GCCG type DE irrespective of
the theory of gravity. The deceleration parameter was also
calculated and plotted against the EoS parameter. From these plots
the notion of an accelerating universe was clearly realized.
Finally some cosmographic parameters, involving different types of
distance measurements have been studied for both the models,
following the power law form of expansion and plot were generated
to characterize the parameters.

\end{document}